\documentclass[11pt,a4paper]{article}
\usepackage[makeroom]{cancel}
\usepackage{bm}
\usepackage{amsmath}
\usepackage{amsfonts}
\usepackage{amssymb}
\usepackage{graphicx}
\usepackage{nicefrac}
\usepackage{authblk}
\usepackage{cite}
\usepackage[left=2.4cm,top=2.4cm,right=2.4cm,bottom=2.4cm,nohead]{geometry}
\DeclareGraphicsExtensions{.png,.jpg,.pdf}
\usepackage{epstopdf}
\usepackage{float}
\usepackage{fouriernc} 
\usepackage{enumitem}
\usepackage[utf8]{inputenc}
\usepackage[english]{babel}
\usepackage{mathtools}
\usepackage{amscd}
\usepackage{mathrsfs}
\usepackage{amsthm}
\usepackage{siunitx}
\usepackage{caption}
\usepackage{subcaption}
\setlist{nolistsep,leftmargin=*}

\usepackage{mhchem} 
\usepackage{cleveref} 
\usepackage{xcolor}
\usepackage{soul}
\usepackage{mdframed}
\usepackage{framed}
\usepackage{lineno}
\usepackage{longtable}
\DeclareMathAlphabet{\mathpzc}{OT1}{pzc}{m}{it}

\title{A concise tutorial review of \\ reverse osmosis and electrodialysis}

\makeatletter

\renewcommand\AB@authnote[1]{\textsuperscript{\normalfont#1}}

\makeatother

\author[1]{P.M. Biesheuvel,}
\author[2]{S. Porada,}
\author[3]{L. Wang,}
\author[4]{R. Wang,}
\author[4]{M. Elimelech,}
\author[5]{J.E. Dykstra}

\affil[1]{Wetsus, European Centre of Excellence for Sustainable Water Technology, Leeuwarden, The~Netherlands.}
\affil[2]{Faculty of Chemistry, 
Wroclaw University of Science and Technology, 
Poland.}
\affil[3]{School of Environmental Science and Engineering, Tongji University, Shanghai, China.} 
\affil[4]{Department of Civil and Environmental Engineering, Rice University, Houston, 
USA.}
\affil[5]{Environmental Technology, Wageningen University, The Netherlands.}

\date{} 

\newcommand{\s}[1]{\mathrm{_{#1}}}

\newcommand{\WR}{\textsc{wr}}

\begin{document}

\maketitle

\vspace{-30pt}

\begin{abstract}

Reverse osmosis (RO) and electrodialysis (ED) are the two most important membrane technologies for water desalination and treatment. Their modes of operation and transport mechanisms are very different, but on a closer look also have many similarities. In this concise version of our tutorial review, we describe state-of-the-art theory for both processes, focusing on simple examples that are helpful for the non-specialist and useful for classroom teaching. Both processes are described by solution-friction (SF) theory which combines ion and water transport across membranes with chemical and mechanical equilibrium at membrane/solution interfaces. We present a derivation of SF theory based on force balances on water and ions and show how the various terms, convection, diffusion, and electromigration, are derived, and how solute partitioning is implemented. Finally, we demonstrate how SF theory accurately describes the osmosis experiment where water and ions are transported in opposite directions across a membrane. 

\end{abstract} 

\vspace{-10pt}
\tableofcontents

\section{Introduction}

\begin{figure}
\centering
\includegraphics[width=1.\textwidth]{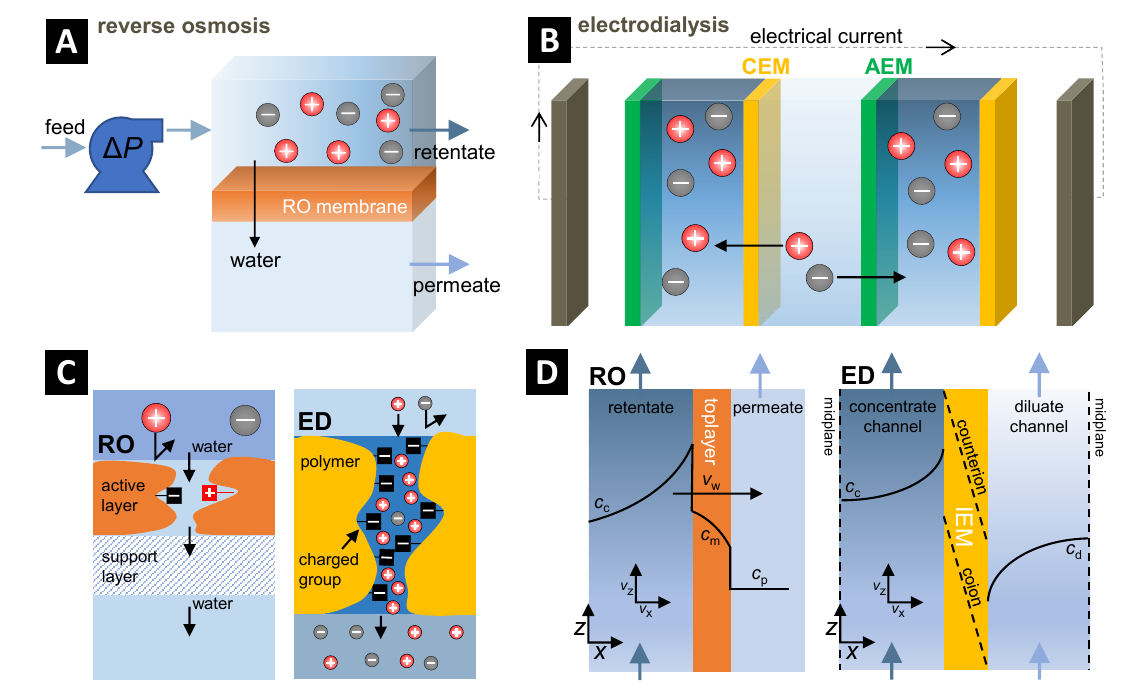}
\vspace{-0mm}
\caption{Overview of reverse osmosis (RO) and electrodialysis (ED). A) In RO, pressure is applied to drive water through a membrane, keeping  most of the solutes on the retentate side. B) In ED, electrical current runs across channels and membranes, resulting in desalination. C) RO and ED membranes have pores through which water and ions move. Because of membrane charge and other effects, solutes partition on the two outsides of a membrane. D) On the upstream side of an RO membrane (retentate side), ion concentrations increase towards the membrane through a concentration polarization layer. On the downstream side, freshwater (permeate) is produced. Concentration profiles develop in an ED cell pair with diluate and concentrate channels separated by ion-exchange membranes.} 
\label{fig_RO_ED_technical_overview}
\end{figure}

Reverse osmosis (RO) and electrodialysis (ED) are the two most applied membrane methodologies for water treatment and 
desalination (deionization)~\cite{Elimelech_2011,Campione_2018}. A brief schematic of both methods is presented in Fig.~\ref{fig_RO_ED_technical_overview}. 
%
%
Water treatment generally refers to the removal of contaminants other than salts, such as organic micropollutants (OMPs), whereas desalination and deionization refer to the removal of salts, thus of ions. RO is a method that uses pressure to drive water through a membrane, keeping  most of the ions and other solutes on the retentate side, producing freshwater as permeate.\footnote{In this work we alternatingly use the words `ion' and `solute' for the charged and uncharged species dissolved in the water.} Nanofiltration (NF) is a companion technology of RO that uses lower pressures, and membranes with larger pore sizes than in RO. 
In NF, the rejection of monovalent ions is much lower than of divalent ions and thus divalent ions can be selectively removed. In ED, water flows through thin channels next to ion-exchange membranes (IEMs) and an applied current pulls the ions from one set of channels through the IEMs to other channels. Though ED and RO are very different and use different physical mechanisms, the underlying transport theory for flow of water and solutes is the same. Thus a generalized treatment is possible that applies to both process types. We also show that key performance indicators on the module level, of relevance for (economic) process optimization, are defined for RO and ED in the same way. 

In this tutorial we focus on desalination of water, i.e., how to use membranes to obtain freshwater from brackish water or seawater by RO and ED. In section~2 we describe thermodynamic equations and metrics that apply to both methods. In sections 3--5 we present the solution-friction (SF) theory, which describes ion and water flow through membranes in combination with the assumption of chemical and mechanical equilibrium at solution/membrane interfaces. In section~3 we use this theory for neutral solutes and present analytical solutions for solute rejection, also including the effect of concentration polarization (CP). For ED we discuss in section~4 a co-current plug flow model also based on SF theory but in the absence of water flow across the membranes. In section~5 we provide a fundamental derivation of SF theory both for ion transport and for water transport, and we subsequently apply SF theory to describe data of an osmosis experiment where we have counter-directional salt and water flow. We also discuss the origin of osmosis, i.e., the reason why water flows from a solution at low salinity to a solution of high salinity. Section 6 is conclusions and outlook.

\begin{figure}
\centering
\includegraphics[width=0.55\textwidth]{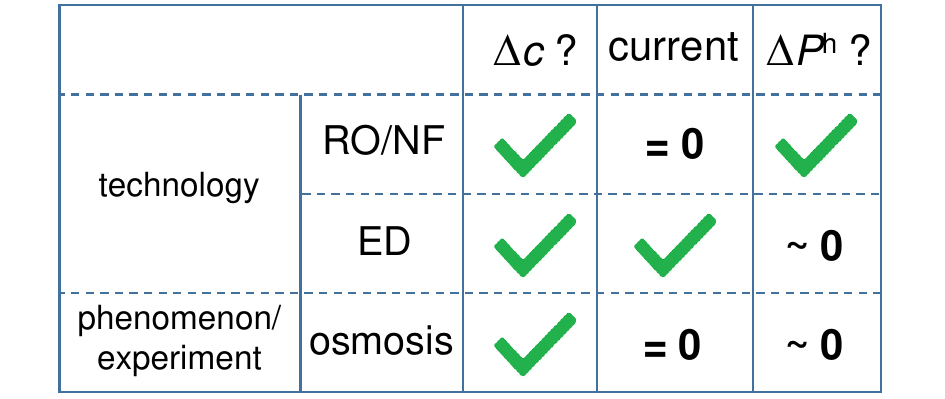}
\vspace{-2mm}
\caption{In reverse osmosis (RO), nanofiltration (NF), and electrodialysis (ED), and in the osmosis experiment, salt concentrations across the membrane are different, there is electrical current, and there is a pressure difference. Even when current is zero, or the pressure difference is zero, or near zero, that information is still an input in the transport models. Thus, on a theoretical level these processes are very similar, and a general theory describes all of them.}
\label{fig_ED_RO_osmosis}
\end{figure}

We illustrate in Fig.~\ref{fig_ED_RO_osmosis} that RO and ED have many aspects in common, 
which they also have in common with the phenomenon of osmosis. In all these processes, the same three features play a key role, as we discuss next. First, in all cases there is a difference in concentration of solutes, or salt, between the two sides of the membrane. Furthermore, there is a certain electrical current, and finally there is a pressure difference across the membrane. In certain cases, the electrical current will be zero, as in RO, and likewise the hydrostatic pressure difference will be (close to) zero in some cases, for instance in the osmosis experiment when the two sides are open to atmosphere and the liquid levels about equal. Nevertheless, in the transport theory, current and pressure being zero is part of the model structure. And thus, when building a theoretical model, it does not matter very much whether current is zero or non-zero, and in all cases a relation for current is part of the model structure. And we have the same situation for the pressure difference (unless in ED we assume water flow across the membrane is zero). 
Because of these similarities in the required theoretical description for RO, ED, and osmosis, a combined discussion of these processes is possible. 

In this tutorial we only address a limited number of topics, and must refer the reader to other literature for many other topics of relevance in RO and ED, such as: for RO, calculation of salt rejection with charged membranes~\cite{Biesheuvel_desalination_2023}, module calculations for RO~\cite{Kimani_2025}, ion selectivity in multi-ionic mixtures~\cite{Starov_1993}, removal of organic micropollutants~\cite{Arola_2019,Verliefde_2007}, module and stack design~\cite{Campione_2018}, fouling abatement~\cite{Lee_2010}, studies of membrane lifetime and robustness, membrane cleaning~\cite{Ghalloussi_2013}, the design of spacers and membrane geometries \cite{Gurreri_2021}, and full system-level cost optimization~\cite{Zhu_2009, Wei_2017, Chehayeb_2018}.

\section{Mass balances, setpoints and metrics, and thermodynamics}

For all water desalination methods, the same overall mass balances apply, the same setpoints and metrics of desalination, and the same thermodynamic equations. These topics are addressed in this section.


\subsection{Overall mass balances for water desalination}

In any water desalination process we start with a feed stream, `f', that is split in a freshwater stream, with index `p' for permeate or product, or `d' for diluate, and a concentrate, retentate, or brine stream, for which index `c' or 'r' is used. We do not address here processes where two different streams are fed into the same desalination unit and salt transfers from one to the other.

Two types of mass balances describe the overall desalination process. First of all we have a volumetric balance, which is  
\begin{equation}
\phi\s{v,f}=\phi\s{v,p}+\phi\s{v,c}
\label{eq_thermo_vol_balance}
\end{equation}
where $\phi_{\text{v},j}$ is a volumetric flow rate, for instance in m\textsuperscript{3}/hr. Secondly there is a mass balance per solute type
\begin{equation}
\phi\s{v,f} \,  c_{\text{f},i}=\phi\s{v,p} \, c_{\text{p},i} + \phi\s{v,c} \, c_{\text{c},i} \, .
\label{eq_thermo_ion_mass_balance}
\end{equation}
where $c_{j,i}$ are concentrations either in mM or M. Instead of describing a process as function of volume flow rates, $\phi_{\text{v},j}$, it can sometimes be helpful to express this problem in terms of volumes, $V_j$. The above balances assume steady-state operation of the desalination system. In the solute balances, we keep track of `conserved species'. For a simple calculation of a 1:1 salt such as NaCl that fully dissociates, we simply follow each ion. but if for instance also some NaCl is present as a salt pair, we track in the balance all species that contain an \ce{Na+}-ion. Similarly, for a problem with \ce{NH3} and \ce{NH4+}, we do not set up a balance in \ce{NH3} only, but a balance in the total ammonia, i.e., in \ce{NH3} plus \ce{NH4^+} jointly~\cite{Biesheuvel_Dykstra_2020}.

\subsection{Setpoints and metrics for water desalination}

Every water desalination process is defined by certain operational characteristics, i.e., objectives, or \textit{setpoints}. These are the amount of water to be treated, the water recovery, and the extent of desalination (or salt removal). Different setpoints can be defined but in the end they describe how much freshwater is obtained from how much feedwater, and what is the salt concentration of the product water. These are considered as setpoints and they must be attained. 

After the setpoints of a process have been defined, \textit{metrics}, or, equivalently, \textit{performance indicators}, come into play. They describe optimized operation. Many metrics are possible. We can for instance define a measure of energy use (energy efficiency), the degree of removal of specific compounds (selectivity), or a measure of how long membranes remain functional. 

\subsubsection{Setpoints for water desalination}

Setpoints for a water desalination process are operational parameters that must be attained, because 
they \textit{define} a certain desalination operation. They are also essential for a calculation of the theoretical minimum energy use (see section~\ref{section_thermo}). 
In practice, the aim is not that these setpoints are reached in an individual desalination unit (module, stack), but they will be the result of operation of a total desalination plant. So a first module desalinates to a certain percentage, and only in a second module the final salt concentration is reached. In that case, setpoints are only reached at the overall level of a full water desalination plant. 

The first setpoint that defines operation of a desalination plant or module, is how much freshwater (diluate, permeate) is obtained per unit time, i.e., the volumetric flowrate of freshwater produced, or permeate, $\phi_{\text{v,p}}$. 

The second setpoint is water recovery, \WR, which is the fraction of the feedwater that is turned into freshwater, thus  
\begin{equation}
\text{\WR}=\frac{\phi_\mathrm{v,p}}{\phi_\mathrm{v,f}} \, .
\end{equation}
Other terminology for water recovery is recovery ratio, and other symbols that are used are \textit{r},  \textit{RR}, or $\alpha$. A related definition is split ratio for the ratio $\phi\s{v,p} {\big / } \phi\s{v,c}$, which is equal to $\WR/\left(1-\WR \right)$.

The third setpoint is the extent of desalination, which one can define as the difference in salt concentration between 
feedwater and product water, $\Delta c = c\s{f} - c\s{p}$. Alternatively, the setpoint can simply be the salt concentration of the freshwater that is obtained. 
Another way to express this setpoint is as a rejection (sometimes called retention), or passage (sometimes called transmission). 
Salt rejection, $R_i$, is defined as
\begin{equation}
R_i = 1 - {c_{\text{p},i}}{\big /} {c_{\text{f},i}} 
\label{eq_rejection}
\end{equation}
while passage, $P_i$, is given by $P_i =1-R_i$. These latter definitions of $R_i$ and $P_i$ are particularly useful for RO, not so much for ED. 
The above three setpoints can be rewritten to other setpoints, but the above formulations are common choices.

In general, we can attain the second and third setpoints in any well-designed RO or ED module, but must then adjust the water flowrate. Besides that, in ED we can tune the current density (or voltage) that is applied to a stack. So in most cases, it will be possible to find operational settings where the second and third setpoints are attained,  for a given flow of water that is treated. The same is valid in RO, where we have flow rates and pressures that can be tuned.  

The third setpoint, the degree of desalination, relates to a total or average desalination of the feedwater as a whole. In reality, most water sources contain many different salts, and we aim for a certain separation level of several critical salts amongst all salts. In practice, still, it is useful to treat the desalination as a setpoint, for instance as a `total $\Delta c$' formulated as a summation over all ions, possibly including a weighing to take into account that some ions are more relevant to remove than others. The degree of desalination for individual ions, related to ion selectivity, is then not a setpoint but treated as a metric.

\subsubsection{Metrics for water desalination}

Metrics, or performance indicators, are those factors that for a certain desalination process, defined by certain setpoints, have values that are as high, or as low, as possible.

A first metric tracks the specific removal of certain key compounds, and the higher the removal of these, the better. So a metric could be a removal degree of one or several key ions, perhaps expressed as a single number by summing over several key ions including a weighing factor describing the importance of the removal of each. 

Another example of a metric is the energy efficiency, $\eta$, which is the ratio of the theoretical minimum energy to achieve a certain desalination (defined by the setpoints) over the actual energy investment. This metric is also called TEE, for thermodynamic energy efficiency. There are many possible definitions for $\eta$, depending on which energy contributions are included in the actual energy input. For instance, for ED we can limit analysis to the electrical energy input required for current to flow across the stack, but we can also include the (much smaller) costs of pumping the water through the channels. In RO we can decide to subtract the recovery of pressure energy. We can take into account efficiencies of power sources and  pumps, and of other electric equipment, for instance to convert (AC) electricity into pumping power for RO, or into direct current for ED~\cite{Werber_2017}. So, a report detailing values of $\eta$ must carefully lay out which energy factors are considered.

Often, instead of an efficiency $\eta$, the inverse factor is reported, which then is the factor by which the actual desalination energy is above the minimum. Generally, values as low as two, for this inverse factor, are considered to be very low, i.e., we  now have an actual energy use that is only a factor two above the thermodynamic minimum. In many cases, especially for water of low salinity, this factor is much higher, for instance 5 or 10, which implies the actual energy is equal to that number of times the thermodynamic minimum.

Other metrics may refer to lifetime of the membranes and track the state of membrane functionality and integrity over time. Another metric is total cost of operation. It is clear that many metrics can be defined and are being used. But important is that setpoints and metrics are clearly defined, both when comparing different materials and operational modes within a certain technology, and when making comparisons between different desalination methods.

\subsection{Minimum energy of desalination}
\label{section_thermo}

For given setpoints, we can calculate the theoretical minimum energy that must be invested to attain that particular degree of desalination. This minimum energy does not depend on what method is used to desalinate. Of course, both in RO and ED there are additional constraints that increase the minimum energy further: for instance, in an RO module we can only apply one pressure, which --on that side of the membrane-- is then effective through the entire module.  And in an ED stack, there is one value of the cell pair voltage, which then applies in the entire stack. All kinds of modifications are possible to relax these system limitations, especially by working with multiple modules that operate at different conditions. But still, the thermodynamic minimum that we analyze in this section, that is system-independent. This minimum only depends on the composition of the water to be treated, temperature, and setpoints. 

This minimum energy that must always be invested to desalinate water, $e\s{min}$, can be calculated on the basis of a thermodynamic analysis of the energy in the three streams that are involved. These three streams are the feedwater `f', the obtained freshwater (permeate, `p', or diluate, `d'), and the concentrate (or, retentate, `c'). In each stream there are multiple ions, each at a concentration $c_i$, and all ions are used in the general equations we will discuss first. For a 1:1 salt of one cation and one anion, we can define a salt concentration, $c_\infty$, which is equal to $c_i$ of each of the two ions involved. 

\subsubsection{The thermodynamics of ideal solutions}

For an ideal solution, the only contribution to the theoretical minimum energy of desalination, $e\s{min}$, is the decrease in ion entropy (free energy of mixing), 
and thus $e\s{min} = - T \Delta S$ where
\begin{equation}
\Delta S = S\s{p} + S\s{c} - S\s{f}
\label{eq_thermo_1}
\end{equation}
with the entropy of each stream given by
\begin{equation}
S_j = - R \cdot \phi_{\text{v},j} \cdot \sum _i  c_{i,j} \ln c_{i,j}
\label{eq_thermo_2}
\end{equation}
where \textit{R} is the gas constant (8.3144 J/mol/K). This equation can be used for any mixture of ions, and irrespective of how effectively each ion is separated in a membrane process. 

A calculation based on these equations is easily performed in spreadsheet software, and results in the input energy in W=J/s required to attain a certain separation in a steady-state process. As mentoned before, it can be helpful to interpret $\phi_{\text{v},j}$ as volumes $V_j$, such that the equation provides the energy in J to treat a certain volume of water. We can define the minimum energy per m\textsuperscript{3} of freshwater produced, $E\s{min}$, as $E\s{min}=e\s{min}/\phi\s{v,p}$. An important result (to check one's calculation) is that ${E}\s{min}\!=\!1.0$ kWh per m\textsuperscript{3} freshwater produced for a 1:1 salt, with 50\% water recovery, i.e., $\text{\WR}\!=\!0.50$, a feed salt concentration $c_{\infty}\!=\!525$~mM, $c_{\infty,\text{p}}\!=\!0$ mM, and temperature $T\!=\!298$ K. Note that a 1:1 salt solution with salt concentration $c_\infty$ contains both cations and anions at that concentration, and 
both ions must be counted. The ion's charge does not play a role in calculating the entropy of desalination.

\subsubsection{
Extensions for non-ideal solutions}
\label{section_non_ideal}

We will discuss various simplifications in a later section, but let us first briefly point out how additional energy terms can be included. Though these additions have a clear physical origin, nevertheless, historically, it is customary to say that we now describe `non-ideal' solutions.

The first non-ideal effect is that of Coulombic interactions between ions, which for a 1:1~salt can be described by the extended Bjerrum equation~\cite{Biesheuvel_2020,Wang_Biesheuvel_arxiv_2024}
\begin{equation}
{f_{\text{cou}}}= - {2 RT c_\infty} \, \left( \tfrac{3}{4} b {c_\infty}^{1/3} + \tfrac{3}{20} b^2  {c_\infty}^{2/3} - 3 b^3 q {c_\infty} \right)
\label{eq_Bjerrum_Maarten2}
\end{equation}
where $f\s{cou}$ is the contribution to the free energy density of a solution, $b\! = \! 0.0605$~mM$^{-1/3}$ and $q$ is a size factor that for \ce{NaCl} is $q \! = \! 0.19$, and $q\!=\!0.125$ for KCl. This contribution, $f_{\text{cou},j}$, can be evaluated for each of the three streams, and then multiplied by the respective volume flow rate, $\phi_{\text{v},j}$, and all terms are added to $e\s{min}$ (with an additional minus-sign for the feed stream). For salt mixtures other than 1:1 salts, numerical analysis of the extended Bjerrum theory is required~\cite{Biesheuvel_2020}.

Another contribution to the energy is an ion volume (size) effect that shows up at higher concentrations, which describes that the (hydrated) ions cannot overlap, i.e., they exclude space for one another. The Carnahan-Starling equation of state can be used to describe this effect (when we assume that all ions have the same hydrated size and all are more or less spherical), and this leads to a further contribution to the free energy, thus to $e\s{min}$. For each flow this contribution is~\cite{Soestbergen_2007} 
\begin{equation}
{f_{\text{vol}}} = { RT  c_{\text{tot}} } \cdot \left( 4 \phi- 3\phi^2 \right) / \left(1-\phi\right)^2  = { RT  c_{\text{tot}} } \cdot \left(  4 \phi + 5 \phi^2 + 6 \phi^3 +  \dots \right) 
\label{eq_CS_ions_free_energy}
\end{equation}
where $c_{\text{tot}}$ is the total ion concentration (for a $z$:$z$ salt, $c\s{tot}\!=\!2 c_\infty$) and $\phi$ is the total volume fraction of all ions, i.e., $\phi\!=\! c\s{tot} \nu\s{i}$, where $\nu\s{i}$ is the molar volume of an ion (with all ions assumed to have the same volume). For each stream, the energy $f_{\text{vol},j}$ can be multiplied by $\phi_{\text{v},j}$, and each contribution can be included in $e\s{min}$ (with a minus-sign for the feed stream). 

As a numerical example, for a 1:1 salt at $c_\infty \!=\! 525$~mM, with ions of a hydrated size of $\sigma \!=\!0.5$~nm (size is twice the radius), the ion volume fraction is $\phi \!= \! 0.0414$. With the Coulombic and volumetric contributions implemented, then for the previous numerical example of a 50\% water recovery, the minimum energy consumption increases from $E\s{min}$=1.00~kWh/m\textsuperscript{3} to  
$E\s{min}$=1.23~kWh/m\textsuperscript{3}. Thus, with the corrections included, the minimum energy increases, and thus for a given achieved energy use in practice, the efficiency is better than in case the ideal value for $E\s{min}$ was assumed.  
Eq.~\eqref{eq_CS_ions_free_energy} can be extended to describe a salt solution where ions have different sizes. Further extensions relate to the energies of 
ion-ion association and ion (de-)protonation, with pK-values of these reactions dependent on salt concentration, but this is more complicated. 

\subsubsection{
Simplifications of the thermodynamic expressions}

Various simplifications are possible to the above general expressions for the theoretical minimum energy. Here we consider a symmetric $z$:$z$ salt solution, with one type of cation and one type of anion. We only discuss the ideal, entropic, contribution. We present results for the energy per m\textsuperscript{3} of permeate (diluate, freshwater) produced. Each of the three streams, feed, permeate and concentrate, is described by a certain salt concentration $c_{j}$ (from this point onward, index `$\infty$' is left out when describing a salt concentration).  
%
%
 
%
%

When we combine the various balances with the expressions for $S_i$, we obtain for $E\s{min}$ the result that\cite{Wang_2020} 
\begin{equation}
\frac{E\s{min}}{2RT} =  \frac{c\s{f}}{\text{\WR}} \ln \frac{c\s{c}}{c\s{f}} - c\s{p} \ln \frac{c\s{c}}{c\s{p}}
\label{eq_thermo_3}
\end{equation}
%
where all concentrations $c_i$ are salt concentrations in mol/m\textsuperscript{3}. 

When the permeate concentration, $c\s{p}$, is zero, i.e., we have complete salt removal, then Eq.~\eqref{eq_thermo_3} simplifies to~\cite{Wang_2020} 
%
\begin{equation}
\frac{E\s{min}}{2 \, c\s{f}  \, RT } = - \frac{\ln \left( 1-\text{\WR}\right)}{\text{\WR}} = 1+\frac{1}{2}\, \text{\WR} + \frac{1}{3}\, \text{\WR}^2 + \dots 
\label{eq_thermo_4}
\end{equation}
%
which shows that the minimum required energy goes up when we aim for a higher water recovery. Note that Eq.~\eqref{eq_thermo_4} is only correct when $c\s{p} = 0$. For non-zero values of $c\s{p}$, this energy, for a very small $\text{\WR}$, is given by
\begin{equation}
\frac{ E\s{min} }{ 2 \, RT}= c\s{f} - c\s{p} - c\s{p} \ln \frac{c\s{f}}{c\s{p}}  \, . 
\label{eq_thermo_5}
\end{equation}
%
The factor 2 on the left of Eqs.~\eqref{eq_thermo_3}--\eqref{eq_thermo_5} is because we consider here a $z$:$z$ salt. For a single neutral solute, this factor is omitted.

\subsection{
Practical energy use in RO and ED}

In the above section the thermodynamic, minimum, energy of a desalination process was discussed. But what is the actual, real, or practical, energy consumption? The main factors are the pressure required to pump fluid through channels, and in RO to press water through the membrane, while in ED we need energy to generate a flow of electrical current across the ED stack.

The energy required to pressurize a stream is $\Delta P^\text{h}$ times flowrate $\phi_{\text{v}}$, which is a number with unit J/s=W. This energy we need to pump feedwater into a module, and to pump water exiting one module into the next one, if it operates at a higher pressure. At the end of a series of modules, it is possible to install an energy recovery device (ERD). If the water on the retentate side (the concentrate; a smaller volume than the flow of feedwater) passes an ERD, exiting at a lower pressure, energy recovery is possible, given by $\eta_\s{ERD}\cdot\Delta P^\text{h} \cdot \phi_{\text{v,c}}$, with $\eta_\s{ERD}$ the energy efficiency of the ERD. 

Another practical limitation, even for ideal operation, is that we generally have a limited number of modules. In RO with a single module, for a membrane that perfectly retains all solutes, the minimum pressure is given by the retentate (concentrate) concentration, which in case of perfect solute rejection relates to $c\s{f}$ by $c\s{c}=c\s{f}/\left(1-\text{\WR}\right)$. So the minimum energy per unit permeate volume, practically, cannot be less than $E\s{min,prac}=2 \,c\s{f}\, RT /\left(1-\text{\WR}\right)$~\cite{Wang_2020}, 
and with $E\s{min}$ given by Eq.~\eqref{eq_thermo_4}, this implies a practical single stage RO efficiency of
\begin{equation}
\eta\s{RO}= - \frac{1-\text{\WR}}{\text{\WR}} \cdot \ln\left(1-\text{\WR}\right) = 1-\nicefrac{1}{2}\cdot\text{\WR}- \nicefrac{1}{6}\cdot\text{\WR}^2-\dots  
\label{eq_some_RO_efficiency}
\end{equation} 

\noindent 
For many reasons, this efficiency cannot be achieved. One reason is that the salt concentration at the membrane is higher than in the bulk of the channel, i.e., the effect of concentration polarization, leading to an over-pressurization factor that reduces efficiency. And to have flow of water across the membrane, we must push harder than the pressure assumed in a calculation of the thermodynamic minimum.

%

Just as in RO, also in ED energy is required to push water through tubings, and 
channels of the ED stack, but this is low compared to the main energy input, which is electricity for running current across the stack. For a certain current \textit{I} (in A, i.e., Ampère), and cell pair voltage $V\s{cp}$, then for each cell pair the electrical energy input is $I \cdot V\s{cp}$ (unit $\text{J}/\text{s}\!=\!\text{W}$), and with \textit{N} cell pairs in an ED stack, this energy is multiplied by \textit{N} to obtain the electrical energy input for the stack. Inside each cell pair the voltage $V\s{cp}$ is due to resistances for current to cross the diluate and concentrate channels, to cross the membranes, and due to the Donnan potentials at the membrane-solution interfaces. These Donnan potentials 
are not a resistance (they are not a direct function of the current), but are a consequence of the thermodynamics of desalination, i.e., the Donnan potential cannot be reduced to zero by running at low currents. Of the Ohmic resistances, the largest loss is in 
the low-salinity diluate channel. 
In an ED stack (of 10s or 100s of cell pairs) there are also two end-compartments in which electronic current transfers to ionic current, and vice-versa~\cite{Biesheuvel_Dykstra_2021_Intro}. There is a voltage loss associated with the Faradaic reactions there, but 
with increasing numbers \textit{N} of cell pairs, this loss becomes less and less important for the energy requirements of a full ED stack. 

The energy thus invested in RO and ED by pressure and electricity (including energy recovery), is always larger than the thermodynamic energy as calculated in section~\ref{section_thermo}. Thus, the ratio of thermodynamic energy \textit{e}\textsubscript{min} over the real, or practical, energy, 
i.e., the energy efficiency $\eta$, 
is always less than one. 

\section{Theory of reverse osmosis with neutral solutes}
\label{section_RO}

\subsection{Flowrates and rejection}

Reverse osmosis (RO) is a process to remove solutes from water by pushing water through a membrane which largely blocks passage for solutes. This is the general mechanism of all pressure-driven membrane processes, 
but for water desalination the size of the pores (free volume) in the membrane must be very small, typically below 1 nm in the selective toplayer of an RO membrane. In classical literature, RO is called hyperfiltration, but nowadays this term has been replaced by RO. Nanofiltration (NF) is similar to RO but has slightly larger pores, and thus the applied hydrostatic pressure is lower, with the rejection of monovalent ions significantly reduced. NF is therefore considered to be suitable for the selective separation of divalent ions and other larger molecules from monovalent salts.

Because RO and NF can be used not only to desalinate water but also to remove other (larger) molecules that may not necessarily be charged, it is useful to start with theory for uncharged solutes. From this point onward, we focus on RO. 
%
%
Based on the SF model, see section~\ref{section_fundamentals}, we can derive for the molar flux of neutral species across a porous membrane (see Eqs.~(1),~(18), and~(19) in ref.~\cite{Kedem_2008}) 
\begin{equation}
J_i=K_{\text{f},i} \, c_{\text{m},i}  J\s{w} - K_{\text{f},i} D_{\text{m},i} \frac{\partial c_{\text{m},i}}{\partial x} 
\label{eq_RO_1}
\end{equation}
where the transmembrane solute flux $J_i$ (in mol/m\textsuperscript{2}/s) and water flux $J\s{w}$ (in m/s) are defined per unit geometric membrane area, thus they are superficial velocities, i.e., they are not pore-based, interstitial, velocities. Concentrations $c_{\text{m},i}$ in the membrane are defined per unit volume of the water-filled pores in the membrane. Eq.~\eqref{eq_RO_1} includes solute transport due to convection, which is the transport of solutes because they flow with, i.e., are convected with, the water that flows through the membrane (first term), and transport because of a solute concentration gradient (second term). Here, \textit{x} is a coordinate axis directed across the membrane. The diffusional term depends on a membrane-based diffusion coefficient, $D_{\text{m},i}$, which includes the effects of porosity and tortuosity. 
The friction factor $K_{\text{f},i}$ depends on the extent of solute-membrane friction, and thus relates to membrane pore size and the (hydrated) size of the solutes. In the absence of solute-membrane friction, the solutes move freely with the water in the pores, while they are still subject to diffusional forces (and electrical forces when charged), and in that case $K_{\text{f},i} \! = \! 1$. However, in an RO process, it is much more likely that $K_{\text{f},i} \ll 1$.


Before continuing with a discussion of transport models, we must first introduce two other model elements. First we describe how the solute concentration just in the membrane is related to that just outside. This condition holds on both sides of a membrane. To that end we make use of a partitioning equation \cite{Starov_1993,Singh_2021,Bowen_2002, Biesheuvel_arxiv_2022,Wang_Biesheuvel_arxiv_2024}, which follows from chemical equilibrium of a species \textit{i} across an interface, resulting for neutral solutes in
\begin{equation}
\Phi_i 
=\frac{c_{\text{m},i}}{c_{\infty,i}}
\label{eq_RO_part}
\end{equation}
where $\Phi$ is a partition coefficient; index `$\text{m}$' refers to a position just in the membrane and `$\infty$' to a position just outside the membrane. When the solutes are charged, i.e., they are ions, an additional Donnan effect arises, which in this paper we treat separately, i.e., do not include in 
$\Phi_i$, see section~\ref{section-RO-ions}. 
The partition coefficient, $\Phi_i$, 
can be due to the chemical affinity of a solute with a certain phase (medium) relative to its affinity with another phase, but also other effects of solute-membrane interaction can be absorbed in $\Phi_i$~\cite{Singh_2021,Biesheuvel_arxiv_2022,Wang_Biesheuvel_arxiv_2024}. In RO with neutral solutes, we always have at least $\Phi_i \! < \! 1$ or $K_{\text{f},i} \!< \! 1$, and likely both factors are $< \! 1$. 

A further element in RO modeling is the permeate equation (Eq.~(8) in~\cite{Starov_1993};~Eq.~(24) in~\cite{Spiegler_1966})
\begin{equation}
c_{\text{p},i} = \frac{J_i }{ J\s{w}  }
\label{eq_RO_permeate}
\end{equation}
%
%
which predicts the concentration of solutes on the permeate side. It a cocurrent flow geometry, it is only valid when at each point in the module the concentration on the permeate side is the same. Thus it is not generally valid when the permeate water flows along the membrane to a single exit point. But for a module in the limit of a very small water recovery, or when on the upstream side we have a mixed solution (stirred tank), it is valid.  
%
In any module it describes the concentration in the exit of the permeate flow, when we use average values for $J_i$ and $J\s{w}$ (each separately averaged over the complete module). 
It is also valid in cross-current flow for certain conditions~\cite{Biesheuvel_Dykstra_2020}. 

We can integrate of Eq.~\eqref{eq_RO_1} across the membrane, and implement Eq.~\eqref{eq_RO_part}, which results in
%
%
%
%
%
\begin{equation}
{J_i}  = \left(1-\sigma_i\right) \, J\s{w} \, \frac{ c_{\text{f},i} \cdot e^{ \text{Pe}_i  } - c_{\text{p},i} } { e^{ \text{Pe}_i  } -1 }
\label{eq_RO_2}
\end{equation}
%
%
where $\text{Pe}_i$ is the membrane Péclet number, $\text{Pe}_i = J\s{w}  / k_{\text{m},i}$, where $k_{\text{m},i}$ is the membrane mass transfer coefficient, given by $k_{\text{m},i} =D_{\text{m},i} {\big /}   L\s{m}$, with \textit{L}\textsubscript{m} membrane thickness. 
In Eq.~\eqref{eq_RO_2}, $\sigma_i$ is the sieving coefficient (also called reflection coefficient) which relates to $K_{\text{f},i}$ and $\Phi_i$ according to 
\begin{equation}
\sigma_i = 1-K_{\text{f},i}\Phi_i \, .
\end{equation}
Eq.~\eqref{eq_RO_2} is the general result for rejection in an RO membrane with neutral solutes~\cite{Spiegler_1966,Kedem_1961,Manning_1968,Wang_1997,Yaroshchuk_2017,Starov_1993}. 
%
%
Calculation results based on this equation are presented in Fig.~\ref{fig_RO_rejection_CP}. Note that when concentration polarization (CP) is also considered, Eq.~\eqref{eq_RO_2} is still valid, but we must replace $c_{\text{f},i}$ by the concentration right at the membrane interface, often written as $c_{\text{int},i}$. 

We combine Eqs.~\eqref{eq_rejection},~\eqref{eq_RO_permeate}, and~\eqref{eq_RO_2}, to obtain for solute rejection
%
%
%
\begin{equation}
R_i= \frac{\left(1- \exp \left( - \text{Pe}_i \right) \right) \cdot \sigma_i}{1-\exp \left( - \text{Pe}_i \right) \cdot \sigma_i} 
\label{eq_RO_7a}
\end{equation}
%
which is 
equivalent to Eq.~(4) in ref.~\cite{Kedem_2008}, Eq.~(3) in ref.~\cite{Wang_1997},  Eq.~(38) in ref.~\cite{Yaroshchuk_2019}, and Eq.~(129) in ref.~\cite{Chmiel_2005}, when we implement that 
$\omega=\left(1-\sigma_i\right)\,k_{\text{m},i}$, $P = P\s{s}= K_\text{f}\Phi_i k_{\text{m},i}$, and thus $k_{\text{m},i}=P\s{s}/\left(1-\sigma_i\right)$. A similar formulation is Eq.~(9) in ref.~\cite{Sutzkover_2000}. 
Note that while Eq.~\eqref{eq_RO_2} is generally valid, at each position in an RO module, Eq.~\eqref{eq_RO_7a} makes use of the permeate equation, Eq.~\eqref{eq_RO_permeate}. All equations in this section are derived for neutral solutes.

%
%

Based on Eq.~\eqref{eq_RO_7a}, rejection in the limit of low transmembrane water flux $J\s{w}$, is given by
\begin{equation}
R_i = \frac{\sigma_i}{ 1-\sigma_i}\; \text{Pe}_i = \frac{\sigma_i}{\omega} \; J\s{w} 
\label{eq_RO_SF_low_Pe}
\end{equation}
which shows that for $J\s{w} \rightarrow 0$, solute rejection goes to zero, and it also shows that to have a non-zero rejection we need $\sigma_i > 0$, and thus either $K_{\mathrm{f},i}$ must be $<1$, or $\Phi_i$ must be $<1$. 

In the other limit, of a high permeate water flux, $J\s{w}$, Eq.~\eqref{eq_RO_2} simplifies to 
\begin{equation}
J_i= K_{\text{f},i}\Phi_i  \,  c_{\text{f},i} \, J\s{w} = \left(1-\sigma_i\right) \,  c_{\text{f},i} \, J\s{w} 
\label{eq_RO_7ab}
\end{equation}
which implies that in this limit, solute flux is only due to convection, and no longer depends on diffusion. 
In this high-Pe limit, rejection is given by 
\begin{equation}
R_i = \sigma_i
\label{eq_RO_limiting_rejection}
\end{equation}
which is Eq.~(30) in ref.~\cite{Spiegler_1966} and Eq.~(34) in ref.~\cite{Chmiel_2005}. Eq.~\eqref{eq_RO_limiting_rejection} describes that an RO membrane has a natural limit in what rejection it can achieve, determined by the extent to which solutes are excluded from the membrane (which implies a value of $\Phi_i \! < \! 1$), and by the extent of solute-membrane friction (which leads to $K_{\text{f},i} \! < \! 1$). If a membrane does not do either, i.e., it does not exclude 
solutes, i.e., $\Phi_i \! = \! 1$, and it does not impose a frictional force on solutes 
at all, i.e., $K_{\text{f},i}=1$, 
then rejection will be zero. Thus, 
for an RO membrane to function, either solutes must be excluded from the membrane, leading to $\Phi_i \!< \! 1$, or there must be a solute-membrane friction, i.e., $K_{\text{f},i} \! < \! 1$, and ideally both.
%
%
~


\subsection{Theory of RO including concentration polarization}

Next we extend the SF-model with concentration polarization. We can envision this effect as if there is a stagnant layer on the upstream side, in front of the membrane~\cite{Bhattacharjee_2001,Biesheuvel_arxiv_2024_CP}. This layer has a certain thickness and all water and solutes must travel across this layer before reaching the membrane. Eq.~\eqref{eq_RO_2}, which describes solute convection and diffusion in a membrane, also applies to the CP layer, now with $\sigma_i \!=\! 0 $, because we are outside the membrane. 
We then arrive at the exponential law for the CP layer~\cite{Biesheuvel_arxiv_2024_CP}
\begin{equation}
c_{\text{int},i} = \left( c_{\text{f},i} - \frac{J_i}{J\s{w}}\right) \cdot \exp \left( \frac{J\s{w} }{k_{\text{cp},i}} \right) + \frac{J_i}{J\s{w}}  
\label{eq_RO_upstream_cp_flux}
\end{equation}
where 
$k_{\text{cp},i}$ is the mass transfer coefficient in the CP layer. 
With more stirring, or a higher crossflow velocity (flux of the water pumped along the membrane),  $k_{\text{cp},i}$ goes up. In Eq.~\eqref{eq_RO_upstream_cp_flux} we use `f' for feed, but that is only correct in an experiment at very low water recovery. More generally, when we use this equation in a 2D module calculation, with \textit{z} a coordinate along the membrane, we should use here the \textit{z}-dependent concentration in the bulk of the channel~\cite{Kimani_2025}. The concentration at the interface between CP layer and membrane (but still in solution) is $c_{\text{int},i}$. 

%

In case Eq.~\eqref{eq_RO_permeate} applies, we can rewrite Eq.~\eqref{eq_RO_upstream_cp_flux} to
\begin{equation}
c_{\text{int},i} = c_{\text{p},i} + \left( c_{\text{f},i}  - c_{\text{p},i}  \right) \exp \left( \frac{J\s{w}}{k_{\text{cp},i}}\right)
\label{eq_RO_upstream_cp}
\end{equation}
and if $c_{\text{int},i}\gg c_{\text{p},i}$, then Eq.~\eqref{eq_RO_upstream_cp} simplifies to $c_{\text{int},i} = c_{\text{f},i} \cdot \exp \left(J\s{w} / k_{\text{cp},i}\right) $. This result is always valid when the membrane blocks all salts or solutes, because then $c_{\text{p},i} \! = \! 0$. 
For any asymmetric binary salt, Eqs.~\eqref{eq_RO_upstream_cp_flux} and  \eqref{eq_RO_upstream_cp} also apply, but the diffusion coefficient $D_{\infty,i}$ is replaced by the harmonic mean diffusion coefficient of the binary salt, $D\s{hm}$~\cite{Biesheuvel_2020}.

\begin{figure}
\centering
\includegraphics[width=0.9\textwidth]{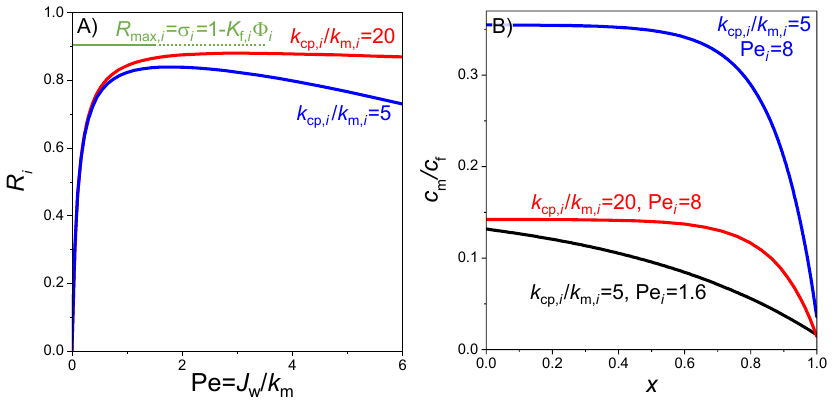}
\caption{Results of calculations for rejection of neutral solutes by an RO membrane, combining membrane transport modeling with solute partitioning and concentration polarization, based on Eqs.~\eqref{eq_RO_7a} and \eqref{eq_RO_limiting_rejection}. A) Rejection as function of permeate water flux, $J\s{w}$, and solute mass transfer coefficients ($\sigma_i\!=\!0.9$). B) Examples of solute concentration profiles across the membrane. All calculations make use of Eq.~\eqref{eq_RO_permeate}.} 
\label{fig_RO_rejection_CP}
\end{figure}

When we combine all these equations, we obtain the following generalized RO solute rejection equation that includes the effect of concentration polarization   
\begin{equation}
R_i= \frac{\left(1- \exp \left( - \text{Pe}_i \right) \right)\, \sigma_i}{\exp\left(J\s{w} / k_{\text{cp},i}\right) \, \left(1-\sigma_i \right)+\left(1- \exp \left( - \text{Pe}_i \right) \right)\,\sigma_i} 
\label{eq_RO_with_cp}
\end{equation}
which is Eq.~(13) in ref.~\cite{Starov_1993}. Eq.~\eqref{eq_RO_with_cp} makes use of Eq.~\eqref{eq_RO_permeate}. When the mass transfer coefficient of the CP layer is large enough, for instance because of sufficient stirring, then Eq.~\eqref{eq_RO_with_cp} simplifies to Eq.~\eqref{eq_RO_7a}. Eq.~\eqref{eq_RO_with_cp} predicts a maximum in rejection, as also shown in Fig.~\ref{fig_RO_rejection_CP}A, and this maximum rejection is at a membrane Pe-number of $\text{Pe}_i=\ln\left(1+k_{\text{cp},i}/k_{\text{m},i}\right)$~(Eq.~(15) in ref.~\cite{Starov_1993}). 



We present in Fig.~\ref{fig_RO_rejection_CP}A results for rejection as described by Eq.~\eqref{eq_RO_with_cp} for various values of 
the ratio $k_{\text{cp},i}/k_{\text{m},i}$. Also results are presented for the limiting rejection given by Eq.~\eqref{eq_RO_limiting_rejection}. For selected conditions, concentration profiles are presented in Fig.~\ref{fig_RO_rejection_CP}B. These profiles can be quite non-linear with a steeper decay in concentration near the permeate side. 

\subsection{Pressure and water flow in RO for neutral solutes}

Having discussed how solutes are retained by RO membranes, we next describe 
the relation between the hydrostatic pressure that is applied across an RO membrane, $\Delta P^{\text{h},\infty}$, 
and the resulting permeate water flux, 
$J\s{w}$. 
%
The force balance on the water 
that will be presented in section~\ref{section_fundamentals} describes fluid flow through porous media in the presence of solutes that move through that same porous structure. The theory describes how water flows because of gradients in hydrostatic and osmotic pressure~\cite{Pappenheimer_1953} (the latter due to gradients in total solute concentration), and is slowed down by friction with the membrane matrix and by friction with solutes. When solutes also 
have friction with the membrane (resulting in $K_{\text{f},i}<1$), then the water velocity decreases: 
with increasing solute-membrane friction, solutes become more of an obstacle for water to flow. 
In the opposite limit when solutes just flow along with the water, without solute-membrane friction, they do not hinder water transport and then for neutral solutes this same theory leads to the classical Darcy equation for fluid flow through a porous medium. 
At the edges of the membrane, we have jumps both in osmotic pressure and hydrostatic pressure~\cite{Chmiel_2005,Sonin_1976,Wang_2021}, and it is essential to understand these to predict the direction of water flow, especially in the absence of a significant hydrostatic pressure. These topics are addressed in section~\ref{section_fundamentals}.

Interesting are the changes of pressure across the CP-layer in front of the membrane. The osmotic pressure increases through the CP layer, because it is a direct function of the local solute concentration, which also increases towards the membrane (because the solutes are rejected by the membrane while water flows through). At the same time, the hydrostatic pressure does not change, but instead the externally applied hydrostatic pressure 
is transferred to the membrane surface unchanged. Only there, at the outer surface of the membrane, are there jumps in the hydrostatic pressure and osmotic pressure upon entering the membrane~\cite{Anderson_1974,Wijmans_1985,Bacchin_2017,Wang_2021}.

For neutral solutes, on the outsides of the membrane, there is a jump down in solute concentration upon entering the membrane when solutes are excluded from the membrane, described by $\Phi_i<1$. A first question is then how the osmotic pressure changes across the interface. In this case also the osmotic pressure goes down, from a value $\Pi^\infty$ just outside the membrane to a value $\Pi^\text{m}$ just inside the membrane, with $\Pi^\text{m}<\Pi^\infty$, when $\Phi_i \! < \! 1$ for all solutes. For ideal solutions the osmotic pressure (both inside and outside a membrane) is proportional to the solute concentration according to $\Pi = c \, RT $ (where \textit{c} is a total concentration, i.e., a summation over all solutes), 
and thus the jump in osmotic pressure (counted in the membrane relative to outside) is $\Delta \Pi = \left( c_{\text{m}} - c_{\infty} \right) RT $, where $c_{\infty}$ and $c_\text{m}$ are total concentrations of solute. If we now include the partition equation, we arrive at $\Delta \Pi = - c_{\infty} \left( 1- \Phi_i  \right)   RT $ (assuming that all solutes have the same $\Phi_i$), which is negative. 
This means the osmotic pressure goes down upon entry into the membrane. The magnitude of this change is proportional to the solute concentrations just outside the membrane, and thus $\Delta \Pi$ can be quite significant on the upstream side of a membrane, while on the downstream side the jump in pressure between outside and inside the membrane is much less when the membrane has a high rejection. 

The next step is to find the relationship between this osmotic pressure change across the membrane surface and the hydrostatic pressure change there. Because of mechanical equilibrium across this interface, the total pressure is invariant across each of the membrane interfaces (between just outside and just inside)~\cite{Sasidhar_1981}, and 
therefore at each of the two membrane interfaces we have
\begin{equation}
P^{\text{h},\infty} - \Pi^{\infty} = P^{\text{h,m}}- \Pi^{\text{m}}
\label{eq_RO_mech_eq}
\end{equation}
where as before `$\infty$' refers to just outside the membrane, and `m' to just inside. Eq.~\eqref{eq_RO_mech_eq} implies that 
a step downward in osmotic pressure upon entry into the membrane (in case $\Phi_i<1$) 
has an associated equally large step downward in hydrostatic pressure~(Eq.~(60a) in ref.~\cite{Manning_1968}). This step downward is the most significant on the side of the membrane with the largest solute concentration. 
%

Inside the membrane, the transmembrane water flux follows from a friction balance that we explain in section~\ref{section_fundamentals}, which in a general form is Eq.~\eqref{eq_water_flow_fund_1}. For neutral solutes, this force balance can be rewritten to
\begin{equation}
J\s{w}= - k_\text{w-m}^{\dagger} \left(\frac{1}{RT} \frac{\partial P^\text{h,m}}{\partial \overline{x}}   - \left(1-K_{\text{f},i}\right) \frac{\partial c\s{m}}{\partial \overline{x}}   \right)
\label{eq_RO_waterflow_1a}
\end{equation}
where we assume for all solutes the same $K_{\text{f},i}$. The nondimensional coordinate $\overline{x}$ is given by $\overline{x}=x/L\s{m}$. Here $k_\text{w-m}^{\dagger}$ is 
a water-membrane permeability that also includes the thickness of the membrane, and indirect friction of water with the membrane via the friction of solutes with the membrane. This latter effect disappears when $K_{\text{f},i}\!=\!1$ and then the $\dagger$-index can be dropped. In that case, i.e., when $K_{\text{f},i}=1$, and if the membrane has constant properties throughout, we have a linear pressure gradient in the membrane that results in water flow  (Eq.~(65) in ref.~\cite{Manning_1968}). In general, also when $K_{\text{f},i}\! \neq \!1$, when we assume a constant value for $k_\text{w-m}^{\dagger}$, we can integrate over the membrane thickness, from $\overline{x}\!=\!0$ to $\overline{x}\!=\!1$, implement Eq.~\eqref{eq_RO_mech_eq} for mechanical equilibrium, as well as the relation for solute partitioning, and then obtain the classical result that the transmembrane water flow rate is given by
\begin{equation}
J\s{w} = \tfrac{1}{RT} \, k_\text{w-m}^{\dagger}  \,  \left(\Delta  P^{\text{h},\infty}-\sigma_i \cdot \Delta  \Pi^\infty \right) 
\label{eq_RO_waterflow_3}
\end{equation}
where $\Delta$ describes a value upstream minus downstream. Eq.~\eqref{eq_RO_waterflow_3} is Eq.~(41) in ref.~\cite{Anderson_1974}, and Eq.~(10.21) in ref.~\cite{Katchalsky_1965}. 
Eq.~\eqref{eq_RO_waterflow_3} describes that for water to flow, the applied pressure must compensate the osmotic pressure difference across the membrane, but less so for a 
lower $\sigma_i$. Thus 
if solutes are not very much excluded by the membrane, water will still flow while the hydrostatic pressure can be much lower than the upstream osmotic pressure~\cite{Yaroshchuk_2019}. Note that $\infty$ here refers to just outside the membrane, which on the upstream side is a position between CP layer and membrane. 

\begin{figure}[H]
\centering
\includegraphics[width=.9\textwidth]{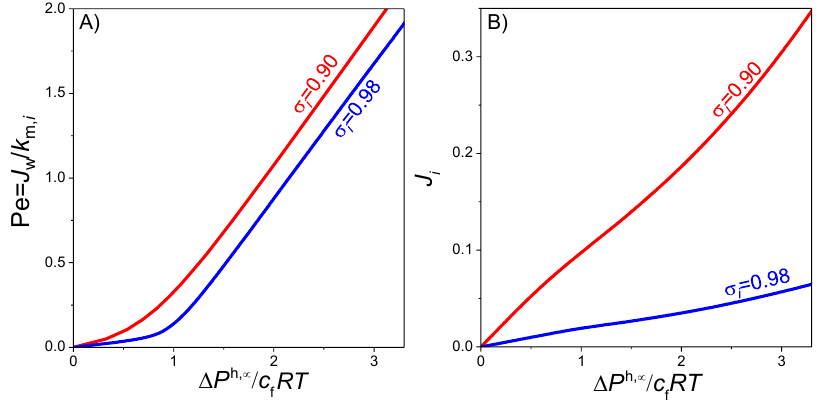}
\caption{Permeate water flux $J\s{w}$ (expressed as a Pe-number) and the molar flux of neutral solutes, $J_i$, as function of $\Delta P^{\text{h},\infty}$, normalized by the feed osmotic pressure, $c\s{f} RT$, across an RO membrane. Results of calculations related to Fig.~\ref{fig_RO_rejection_CP} ($k_{\text{cp},i}/k_{\text{m},i}\!=\!5$, $k_\text{w-m}^{\dagger} \Pi_\text{f} /  RT k_{\text{m},i} = 1$).} 
\label{fig_RO_pressure_Pe_Ji}
\end{figure}

In Fig.~\ref{fig_RO_pressure_Pe_Ji}A, we analyze how the water flux depends on the pressure. The closer the reflection coefficient, $\sigma_i$, is to one, the more clearly there is a  critical pressure that must be overcome before water starts to flow across the membrane. Furthermore, Fig.~\ref{fig_RO_pressure_Pe_Ji}B illustrates that in the SF model the flow of solutes gradually increases with pressure. This concludes our introduction for the theory of RO for neutral solutes. Advanced models for RO for water desalination (i.e., with ions), in one and two dimensions, are discussed in refs.~\cite{Biesheuvel_desalination_2023,Biesheuvel_arxiv_2024_CP,Biesheuvel_Dykstra_2020}. It is also explained there how the osmotic pressure in a salt solution is lower than the ideal pressure because of Coulombic interactions between ions, as was addressed in section~\ref{section_non_ideal}.

\section{Theory of electrodialysis for a 1:1 salt solution}
\label{section_ED}

In this section we describe theory for electrodialysis (ED) in a co-current design, for a 1:1 salt. In ED, feedwater flows into two adjacent sets of channels, see Fig.~\ref{fig_RO_ED_technical_overview}B. In one series 
of channels the water will be desalinated, which are the diluate channels, denoted by `d'. Salt ions move from these d-channels to the concentrate `c' channels. In many practical ED stacks, the direction of flow in a channel is at a 90$^\circ$ angle to the flow in adjacent channels. This is called a cross-flow geometry. This geometry is 
more complicated to model, 
and thus in this section we describe that we have perfect co-current flow, where the two flows enter the module from the same side and flow through the stack in parallel. The two channels can still be different in width and the flowrates (thus residence times) can be different. 

An ED stack consists of many cell pairs, with electrodes on the two sides of the stack, where electronic current becomes ionic current. In this review we do not discuss the electrodes but we focus on the repeating unit of an ED stack, which is the membrane cell pair, see Fig.~\ref{fig_RO_ED_technical_overview}B, which consists of two membranes and two flow channels. As mentioned, in the flow channels we assume co-current flow of the water along the membranes, from inlet to exit of the channel. While the water flows through these thin channels between the membranes, ions move from the d-channels to the c-channels by transport across the membranes. This process is possible because --driven by the electrical current-- anions move in one direction out of each d-channel, to be transported through membranes that are selective for anions. These are anion-exchange membranes (AEMs), which are membranes in which the water-filled pores are lined with high concentrations of positive membrane charges (of the order of 5 M fixed membrane charge per volume of water in the membrane). The cations move in the other direction, passing cation-exchange membranes (CEMs). These CEMs have a high concentration of fixed negative charge, again of the order of 5 M, and they preferentially allow access to cations, largely blocking the passage of anions. Each d-channel has such an AEM on one side, and a CEM on the other.  The net effect of this layout of a cell pair, of a sequence of an AEM, d-channel, CEM, and c-channel, and this repeated tens to hundreds of times, is that the d-channels are being desalinated, while the salinity in the c-channels increases going from entrance to exit of the channel. Thus in the AEMs anions are the main charge carrier, while in the CEMs cations carry most of the charge. These ions in their respective membrane are the counterions (anions in an AEM, cations in a CEM). The minority ions, that ideally are fully blocked, are called coions (cations in AEM, anions in CEM). Though in ED there is also water flowing through the membrane, we will neglect that aspect in this tutorial. Water recovery, \WR, is then directly set by how much of the feedwater flows to the c-channels and how much to the d-channels, namely $\text{\WR}=1/\left(1+\phi\s{v,c}/\phi\s{v,d}\right)$. Thus, for equal flow rates to the two channels, we have $\text{\WR}\!=\!0.50$.

For each channel in the ED cell pair, we set up a steady-state plug flow model, to describe the flow from inlet to exit of a channel, with ion transport through the membranes included, directed sideways. In each channel, at each position $z$, where $z$ is the direction along the membranes, there are concentration profiles across the channel, from membrane to membrane, but they 
can be relatively minor when the channels are narrow.\footnote{This is not always the case, and sometimes these profiles matter a lot. Also with three or more different ions, instead of a simple binary salt, the situation in the flow channels can be markedly different, and concentration profiles across the channel must then be modelled.} Thus in this tutorial, in each channel we use a \textit{z}-dependent mass balance that does not have any dependence on position in 
the perpendicular \textit{x}-direction, which 
is the direction towards the membranes, 
at right angles to the \textit{z}-direction of water flow along the membranes. Then a mass balance for an ion \textit{i} at some position in a flow channel (either a d-channnel or c-channel) is
\begin{equation}
p \frac{\partial c_i}{\partial t} = - v\s{z} \frac{\partial c_i}{\partial z} \pm  a \cdot  \left( J_{\text{AEM},i} + J_{\text{CEM},i}\right) 
\label{eq_ED_1}
\end{equation}
where \textit{p} is the porosity (open volume fraction) of the channel, and $v\s{z}$ the velocity of the water through the channel, in the direction along the membrane. This is an average velocity, i.e., it is a water flowrate in m\textsuperscript{3}/s divided by the cross-sectional area of a channel (the cross-section through which the water flows). It is a superficial velocity, i.e., the presence of a spacer material that reduces \textit{p} does not change $v\s{z}$. The specific membrane surface area is \textit{a}, which is the area of one membrane lining a channel divided by the channel volume. And thus $a=1/L\s{ch}$, where $L\s{ch}$ is the width of the channel. We use the symbol $\pm$ in Eq.~\eqref{eq_ED_1} to avoid technicalities of how to define the directions of fluxes, because that is not essential for this tutorial.

We now consider a binary 1:1 salt. We add up Eq.~\eqref{eq_ED_1} for anions and cations, assume steady-state, thus the term $\partial c_i/ \partial t$ is zero, and consider that at each position \textit{z} we have electroneutrality, i.e., $c_+ \! = \! c_-$, and from now onward we use $c$ for the salt concentration, i.e., $c \! = \! c_{+} \! = \! c_{-}$ which is \textit{z}-dependent, and different between the c- and the d-channels. 
Thus we add up Eq.~\eqref{eq_ED_1} for the two ions, and then divide by 2. The summation of the four fluxes $J_{\text{IEM},i}$ is twice the total salt flux, $J\s{salt}$, which is the flux of salt from d- to c-channels, per unit area of one membrane. We then arrive at
\begin{equation}
 v\s{z} \frac{\partial c_j}{\partial z} = \pm  a_j J\s{salt} = \pm a_j \lambda\s{cp} \frac{I}{F}\ 
\label{eq_ED_2}
\end{equation}
where we introduce the current efficiency $\lambda\s{cp}$, a process parameter that is \textit{z}-dependent in the cell pair, i.e., $\lambda\s{cp}(z)$, and we introduce the current density $I$ in A/m\textsuperscript{2}. 
In an ED stack, in a cell pair, from entrance to exit (i.e., from $z\!=\!0$ to $z\!=\!\ell$, with $\ell$ the length of the channel in flow direction), current density \textit{I} will change with \textit{z}. We cannot make the current constant in a normal ED stack. This is because there are two end-electrodes that have a certain voltage between them. Therefore it is the cell pair voltage, $V\s{cp}$, which will be the same at each \textit{z}-position in the cell, and 
because Donnan potentials and resistances increase with \textit{z}, current will decrease in that direction. 

Eq.~\eqref{eq_ED_2} introduces a cell-pair based current efficiency $\lambda\s{cp}$ which relates to the total salt transport flux between the d- and c-channels, $J\s{salt}$, and the current, \textit{I}, all \textit{z}-dependent. To calculate $\lambda\s{cp}$ and $J\s{salt}$ we must solve at each \textit{z}-coordinate a membrane model for the AEM and the CEM, and in each membrane describe the flow of counterions and coions. 
In such a model all parameters, including the diffusion coefficients of the ions, and the membrane charge, can be different between AEM and CEM. This calculation is possible, but in this review we simplify this situation, and assume the AEM and CEM are each other's perfect mirror image. Thus, the fixed membrane density in an AEM, denoted by the symbol \textit{X}, has the same magnitude as the membrane charge of a CEM, and only the sign of this membrane charge is opposite. Furthermore, the diffusion coefficient of the counterions in the AEM and CEM is the same, and the same for the coions (coions and counterions can still have diffusion coefficients that are different from one another). With these assumptions, 
the cell-pair based $\lambda\s{cp}$ is now the same as a single membrane-based current efficiency $\lambda\s{th}$, where we use index `th' for theoretical, because this efficiency cannot be measured but can only be obtained from a theoretical model of a single ion-exchange membrane. 
Thus, in Eq.~\eqref{eq_ED_2} we make the replacement $\lambda\s{cp} \rightarrow \lambda\s{th}$. 

We combine Eq.~\eqref{eq_ED_2} for the d- and c-channel, and then integrate between inlet and any position \textit{z}, which leads to the overall salt mass balance
\begin{equation}
\phi\s{v,d} \left(c\s{d}(z) - c\s{d,in} \right) + \phi\s{v,c} \left(c\s{c}(z) - c\s{c,in} \right)  = 0
\label{eq_ED_5}
\end{equation}
which expresses that all salt that is removed from one channel ends up in the other channel. This balance assumes there is no water flow across the membranes. If both channels have the same inlet concentration, then $c\s{d,in}=c\s{c,in}=c\s{f}$. We can now use Eq.~\eqref{eq_ED_5} to calculated $c\s{c}$, and only solve Eq.~\eqref{eq_ED_2} for the diluate channel. Next, we introduce the time-on-stream, $t^* = z/v\s{z}$, and rewrite Eq.~\eqref{eq_ED_2} for the d-channel to 
\begin{equation}
\frac{\partial c\s{d}}{\partial t^*} = - a\s{d} \, \lambda\s{th} \, \frac{I}{F} \, .
\label{eq_ED_6}
\end{equation}
and this representation shows that this model can be used for two very different cases: first, the above-discussed geometry of an ED cell pair with two co-current flow channels, but we can also use it for an ED stack with relatively low desalination per pass, where the effluents of each channel are rerouted to two storage tanks from which the same channel is fed again. In this second type of calculation, $t^*$ really is a time. In that case, the area/volume ratio, $a\s{d}$, in Eq.~\eqref{eq_ED_6} is the area of one membrane times the number of cell pairs, divided by the volume of the d-reservoir (including volumes of tubing and the channels in the ED stack). The other modification for this second type of calculation is that in Eq.~\eqref{eq_ED_5}, volume flow rates are replaced by the two reservoir volumes. 
%
%
%
Finally, we have to find expressions for current efficiency, $\lambda\s{th}$, and current density, $I$, as function of $c\s{d}$ 
and the applied cell pair voltage $V\s{cp}$. This is discussed in the next section.

\subsection{Calculation of current and current efficiency in ED}
\label{section_current_current_eff_ED}

To calculate current and current efficiency, we must relate current density \textit{I} to the voltage drops over the membranes and flow channels. The cell pair voltage, $V\s{cp}$, is a summation of the voltage drops across two channels (d- and c-channels), two membranes (AEM and CEM), and in total four Donnan potential drops (at each channel/membrane interface in the cell pair). For a symmetric system, the four Donnan potentials simplify to two times the same difference between two Donnan potentials. In the symmetric model we also assume each membrane has the same resistance. The two channels will, as desalination progresses, have very different resistances. In each membrane, the relationship between current density and voltage can be described by 
\begin{equation}
I= \pm k^*_\text{m} F   |X|  \Delta\phi\s{m}
\label{eq_ED_7}
\end{equation}
where $k^*_\text{m}$ is a membrane transport coefficient, given by $k^*_\text{m}=K_{\text{f},i}k\s{m}$, where $k\s{m}=D_{\text{m},i}/L\s{m}$, with $L\s{m}$ membrane thickness. Furthermore, $|X|$ is the magnitude of the membrane charge density (with unit mM), and $\Delta \phi\s{m}$ is a dimensionless potential drop across the inner part of the membrane, excluding the Donnan potentials at the two surfaces of each membrane. These we discuss further on. Eq.~\eqref{eq_ED_7} assumes that both ions have the same diffusion coefficient in the membrane, $D_{\text{m},i}$, and the same $K_{\text{f},i}$, and thus $k^*_\text{m}$ is the same for both ions. 
Eq.~\eqref{eq_ED_7} also assumes that the membrane charge density, $|X|$, is large compared to salt concentrations outside the membrane. (In a generalized version of Eq.~\eqref{eq_ED_7}, the term $|X|$ is replaced by an average total ions concentration, $\left\langle c\s{T,m} \right\rangle$, but for a good ion exchange membrane these two numbers are very close.) Note that dimensionless potentials, $\phi$, can always be multiplied by the thermal voltage, $V\s{T}=RT/F\sim 25.6$ mV (at room temperature), to arrive at a dimensional voltage in (m)V.

Similar to Eq.~\eqref{eq_ED_7}, we have a current-voltage relationship across the d- and c-channels, given by
\begin{equation}
I=  \pm 2 k_{\text{ch}} F c    \Delta\phi_{\text{ch}}
\label{eq_ED_8}
\end{equation}
where $k_{\text{ch}}$ is a channel transfer coefficient, given by $k_{\text{ch}} = \varepsilon D_\infty / L_{\text{ch}}$, where $D_\infty$ is the mean diffusion coefficient of ions in free solution. The factor $\varepsilon$ is the channel porosity \textit{p} divided by a tortuosity factor $\bm{\tau}$, i.e., $\varepsilon = p / \bm{\tau}$, which describes how in the spacer channel diffusion is slower than in free solution because a spacer structure fills the channel. 
The channel width is $L_{\text{ch}}$. Note the factor 2 in Eq.~\eqref{eq_ED_8} which 
relates to both anions and cations contributing equally to current flow. 
Eq.~\eqref{eq_ED_8} is valid when the two ions have the same diffusion coefficient,\footnote{As long as we have a binary salt, also for ions of different valency and diffusion coefficient we can derive an expression such as Eq.~\eqref{eq_ED_8}, based on the harmonic mean diffusion coefficient.} and it neglect concentration changes across the width of the channel.

Next we consider the four Donnan potentials at the four membrane/solution interfaces. In the symmetric system we discuss, the two Donnan potentials at the AEM are exact mirror-images of those at the CEM, because we assume the same $|X|$ for AEM and CEM, so we only have to discuss one of these membranes here. We furthermore assume that we only have monovalent ions in solution. 
The extended Boltzmann equation for each ion, distributing between just in the membrane and just outside, is
\begin{equation}
c_{\text{m},i} = c_{\infty,i}  \; \Phi_i \;  \exp \left(- z_i \Delta \phi\s{D}   \right)
\label{eq_Donnan_1}
\end{equation}
where $\infty$ refers to a position just outside the membrane (in the ED model \textit{z}-dependent), and Eq.~\eqref{eq_Donnan_1} applies to the interface of the membrane with electrolyte both in the d- and c-channels. The Donnan potential at each of these interfaces is the potential just in the membrane, relative to just outside, in solution. A partition coefficient, $\Phi_i$, is also included, which we discussed in section~\ref{section_RO}. For IEMs, $\Phi_i$ is likely not very small, not $\ll 0.5$.

Electroneutrality in the membrane is given by
\begin{equation}
c_{\text{m},+} - c_{\text{m},-} +  X =0 
\label{eq_Donnan_2}
\end{equation}
where \textit{X} can be positive or negative. Based on Eqs.~\eqref{eq_Donnan_1} and~\eqref{eq_Donnan_2} we can derive the Donnan potential
\begin{equation}
\Delta \phi\s{D} = \sinh^{-1} \frac{X}{2 \Phi_i c_\infty} 
\label{eq_Donnan_3}
\end{equation}
where sinh(\textit{x}) is $\text{\textonehalf} \left(\exp\left(x\right)-\exp\left(-x\right)\right)$, and sinh\textsuperscript{-1} is the inverse function. 
For both ions we assume here the same value of $\Phi_i$. 
As Eq.~\eqref{eq_Donnan_3} shows, for a membrane with positive fixed charge, i.e., an AEM, which needs to `pull in the anions', the Donnan potential is positive, i.e., from just outside to just inside the membrane, the potential goes up.

The eight voltage drops described in the above equations can all be added up (taking care of appropriate choices of $\pm$-signs), to arrive after multiplication with $V\s{T}$, at the cell pair voltage, $V\s{cp}$. And this $V\s{cp}$ is invariant with position \textit{z}; see Fig.~2 in ref.~\cite{Tedesco_2018} for an example of a voltage profile. (Thus the eight individual voltages described above, all change with \textit{z}, but their summation, which is $V\s{cp}$, does not.)

Before we show the result of such a combined expression for $V\s{cp}$ as function of current density, \textit{I}, and salt concentrations, $c\s{d}$ and $c\s{c}$, we first discuss the Donnan potential, Eq.~\eqref{eq_Donnan_3}. This Donnan potential must be evaluated on the two sides of a membrane, resulting in
\begin{equation}
\Delta \phi\s{D,tot} =  \sinh^{-1} \frac{X}{2 \Phi_i c\s{{d}}} - \sinh^{-1} \frac{X}{2 \Phi_i c\s{{c}}} 
\label{eq_Donnan_4}
\end{equation}
which we can call the total Donnan potential across a membrane. (This excludes the membrane potential across the inner part of the membrane due to current, as given by Eq.~\eqref{eq_ED_7}.) From this point forward, we drop index $\infty$ once again. 
Eq.~\eqref{eq_Donnan_4} is not generally used, but 
instead, often found in literature is
\begin{equation}
\pm \Delta \phi\s{D,tot} = \ln \frac{c\s{c}}{c\s{d}} 
\label{eq_Donnan_5}
\end{equation}
and this expression is valid for an ideal membrane, one that perfectly blocks coions. There is no dependence on membrane charge here. A prefactor in front of the ln-term can be added, which is then called the permselectivity, which describes the difference between the ideal ln-term and the measured Donnan potential. The $\pm$-symbol depends on the sign of the membrane charge and on whether the Donnan potential is defined from c-side to d-side, or vice-versa.

Now, interestingly, Eq.~\eqref{eq_Donnan_5} follows from Eq.~\eqref{eq_Donnan_4} as a first term in a Taylor-expansion around the point that $X^{-1} \rightarrow 0$. This first term suffices when on both sides of the membrane the salt concentration is very low compared to \textit{X}. But we can now 
add the second term in the Taylor expansion as well, and obtain a very accurate expression for the Donnan potential, given by
\begin{equation}
\pm \Delta \phi\s{D,tot} = \ln \frac{c\s{c}}{c\s{d}} - \Phi_i^2 \left(c_\text{c}^2-c_\text{d}^2 \right) {\big /} X^2
\label{eq_Donnan_6}
\end{equation}
where the total Donnan potential decreases (in magnitude) when the membrane charge density \textit{X} goes down, or when the difference between $c\s{c}$ and $c\s{d}$ increases. This novel equation is only valid as a first correction to Eq.~\eqref{eq_Donnan_5} when the salt concentrations on both sides are sufficiently lower than $|X|$. 
For higher salt concentrations, we must use Eq.~\eqref{eq_Donnan_4}.  
Note that in Eqs.~\eqref{eq_Donnan_4} and~\eqref{eq_Donnan_6} it is important to use the same unit for \textit{c} and for \textit{X}, either both in M or both in mM, i.e., mol/m\textsuperscript{3}.

Thus, the new expression provides a correction to the Donnan potential for finite values of \textit{X}, without invoking the empirical concept of a permselectivity. The additional term shows that the correction to ideality not only depends on the membrane charge density \textit{X}, but also on $c\s{d}$ and $c\s{c}$, and this clearly highlights how the concept of a permselectivity is not an intrinsic membrane property, but for a given membrane varies with salt concentration, even along the \textit{z}-coordinate within the same ED module (or changes in time in a batch experiment with changing reservoir concentrations).

In general, for multi-ionic salt mixtures, and when we also include the partition coeffcient, $\Phi_i$, possibly different between all ions, it is advisable to return to a Boltzmann equation for each ion, Eq.~\eqref{eq_Donnan_1}, and solve that in combination with electroneutrality in the membrane. In further model extensions, we can include how membrane charge is a function of pH, or of the concentration of other ions, such as the \ce{Ca^2+}-concentration just in the membrane (in case these ions adsorb), 
and these ion concentrations in turn depend on the Donnan potential, thus on \textit{X}. Also other acid-base associations between ions can be included, such as the protonation of an ion, for instance \ce{NH3} that can react to \ce{NH4+}, with the distribution between these two species depending on pK and local pH~\cite{Hwang_1987}. Furthermore, concentration profiles across the width of the channels (especially for solutions with three or more types of ions) result in ion concentrations just outside the membrane (which are the concentrations that are used in the Donnan equilibria) that can be quite different from the channel-averaged concentrations at that \textit{z}-position. Other extensions are transmembrane water flow, which also modifies ion concentrations just next to the membrane. Thus, the simplified model explained in this section typically does not suffice for practical conditions.

For the ideal model that we consider, we can now add up all eight voltages, and obtain one final equation for voltage $V\s{cp}$ as function of current density \textit{I}. The results shown next assume that c- and d-channels have the same width, $L\s{ch}$, and we already discussed the assumption that each membrane has the same charge density, $|{X}|$, and the same transport properties, thus the same $k^*_\text{m}$. We then arrive at
\begin{equation}
\frac{V\s{cp}}{V\s{T}} - 2 \, \left( \ln \frac{c\s{c}}{c\s{d}} - \frac{\Phi_i^2}{X^2} \left(c_\text{c}^2-c_\text{d}^2\right) \right) =\frac{I}{F} \, \, \left( \frac{1}{2\,k\s{ch}} \left(\frac{1}{c\s{d}} + \frac{1}{c\s{c}}  \right) +   \frac{2}{k^*_\text{m} |X|} \right)
\label{eq_ED_IV_1}
\end{equation}
where $c\s{c}$, $c\s{d}$, and \textit{I} are all \textit{z}-dependent. 
%
%

Finally we evaluate the current efficiency $\lambda\s{th}$, at each \textit{z}-position. To that end we must establish what is the flux of counterions through each membrane, 
which is the desired effect, leading to desalination, and the flux of coions, which reduces desalination because coions leak out of the 
c-channels to the d-channels. Based on fluxes through a single membrane, $\lambda\s{th}$ is given by 
\begin{equation}
\lambda\s{th} = \frac{J\s{m,+}+J\s{m,-}}{J\s{m,+}-J\s{m,-}} =  F \;\frac{ J\s{m,ions} }{ I}
\label{eq_ED_lambda_1}
\end{equation}
where $J\s{m,ions}$ is the total ions flow through a membrane, $J\s{m,ions}=J\s{m,+}+J\s{m,-}$, and the current density is $I=F \, \left(J\s{m,+} - J\s{m,-}  \right)$. The factor $\lambda\s{th}$ is only defined for desalination of a 1:1 salt. When we start with both channels fed with the same water, $\lambda\s{th}$ will be between 0 and 1, with $\lambda\s{th}=1$ for ideal operation, with only counterion transport through each membrane, and then for each unit of charge a full salt molecule is removed from a diluate channel. Instead, in the limit of $\lambda\s{th}=0$, current is passing the membrane, but there is no net salt removal. This last situation develops if we make an ED cell very long, or continue to recycle salt back into an ED stack for a very long time.

Next, we analyze ion transport, both of counterions and coions, across a membrane. To that end we use the Nernst-Planck (NP) equation jointly with local electroneutrality. For each ion, the extended NP-equation is
\begin{equation}
J_i = - K_{\text{f},i} D_{\text{m},i} \left( \frac{\partial c_{\text{m},i}}{\partial x} +z_i c_{\text{m},i}\frac{\partial \phi}{\partial x} \right)
\label{eq_ED_NP_1}
\end{equation}
where $\phi$ is the inner-membrane potential. 
If we assume the same diffusion coefficient for anions and cations in the membrane, and then 
subtract anion flux from cation flux, we arrive at
\begin{equation}
I/F = J_+ - J_- = - K_{\text{f},i} D_{\text{m}} c\s{T,m}  \frac{\partial \phi}{\partial x} 
\label{eq_ED_NP_2}
\end{equation}
where $c\s{T,m}$ is the sum of concentrations of the two ions in the membrane, i.e., $c\s{T,m}\! = \! c\s{m,+} \! + \! c\s{m,-}$. We assume the same $K_{\text{f},i}$ for the two ions. 

Note that current density \textit{I} in Eq.~\eqref{eq_ED_NP_2}, calculated based on ion transport in a membrane, is the same current density that also crosses (at that same \textit{z}-position) each of the flow channels, and crosses the other membrane too; i.e., there is (at each \textit{z}-position) only one value of current \textit{I}, the same at each location across each of the channels, and it has the same value in all the other cell pairs (at that \textit{z}-position), because they are all exact copies of one another. 


The NP equation for anions and cations also leads to an expression for the total ions flux
\begin{equation}
J\s{ions,m} = J_+ + J_- = - K_{\text{f},i} D_{\text{m}} \left( \frac{\partial c\s{T,m}}{\partial x} - X  \frac{\partial \phi}{\partial x} \right) \, .
\label{eq_ED_NP_3}
\end{equation}
In the derivation of Eq.~\eqref{eq_ED_NP_2} and~\eqref{eq_ED_NP_3}, the electroneutrality condition in the membrane is implemented. 
Because of steady state, both equations can be integrated across the membrane, leading for current \textit{I} to Eq.~\eqref{eq_ED_7}, and for $J\s{ions,m}$ to 
\begin{equation}
J\s{ions,m} = \pm k^*_\text{m} \left(\Delta c\s{T,m} - X \Delta \phi\s{m} \right) 
\label{eq_ED_Jionsm}
\end{equation}
where $\Delta c\s{T,m} $ is the difference between $c\s{T,m}$ at the two outsides of the membrane, which we calculate based on Eqs.~\eqref{eq_Donnan_1} and~\eqref{eq_Donnan_2}. For a 1:1 salt, we arrive at
\begin{equation}
c_{\text{T,m}}=\sqrt{X^2+ \left( 2 \Phi_i c_{\infty}\right)^2} = |X| + 2 \, \Phi_i^2 c_{\infty}^2 \, {\big /} \, \left|X \right| \, + \, \dots 
\label{eq_ED_cTm}
\end{equation}
%

%
%
We can now calculate $\lambda\s{th}$ according to Eq.~\eqref{eq_ED_lambda_1} by implementing Eqs.~\eqref{eq_ED_7} and~\eqref{eq_ED_Jionsm}, and obtain
\begin{equation}
\lambda\s{th}= 1 - {k^*_\text{m} F \Delta c\s{T,m}}\, {\big /} \,{ |I|} \, .
\label{eq_ED_lambda_th_1}
\end{equation}
%
%
At high enough $|X|$ (relative to $c\s{c}$ and $c\s{d}$), we can derive from Eq.~\eqref{eq_ED_cTm} that the difference in $c\s{T,m}$ across the membrane can be simplified to $\Delta c\s{T,m} = 2 \, \Phi_i^2 \,  \left( c_\text{c}^2 - c_\text{d}^2 \right) {\big /} \, |X| $, 
a result that we can implement in Eq.~\eqref{eq_ED_lambda_th_1} to arrive at 
\begin{equation}
\lambda\s{th}= 1 - \frac{ 2 k^*_\text{m} F }{|X| |I| } \, \Phi_i^2 \, \left(\vphantom{\frac{ 2 k F }{|X| \,|I| }}  c_\text{c}^2 - c_\text{d}^2 \right)  
\label{eq_ED_lambda_4}
\end{equation}
which is the final equation that we need to solve the full co-current ED model. All fluxes and concentrations in Eq.~\eqref{eq_ED_Jionsm} depend on position \textit{z} in the module, or alternatively, depend on time, or time-on-stream, $t^*$. 

Eq.~\eqref{eq_ED_lambda_4} shows that $\lambda\s{th}$ will be high, near unity, at high currents and a high membrane charge, and for low salt concentration differences, but it will go down for lower currents and lower membrane charge. 
It also goes down when $k^*_\text{m}$ goes up, for instance because membranes are thin. Indeed, for very thin membranes, ED does not work, because then there is a large flux of coions across the membranes and that limits the desalination process~\cite{Tedesco_2018}. 
Eq.~\eqref{eq_ED_lambda_4} illustrates that $\lambda\s{th}$ strongly depends on position in the cell, because current density \textit{I} changes with \textit{z}, as well as concentrations $c\s{d}$ and $c\s{c}$.

\begin{figure}[H]
\centering
\includegraphics[width=1.0\textwidth]{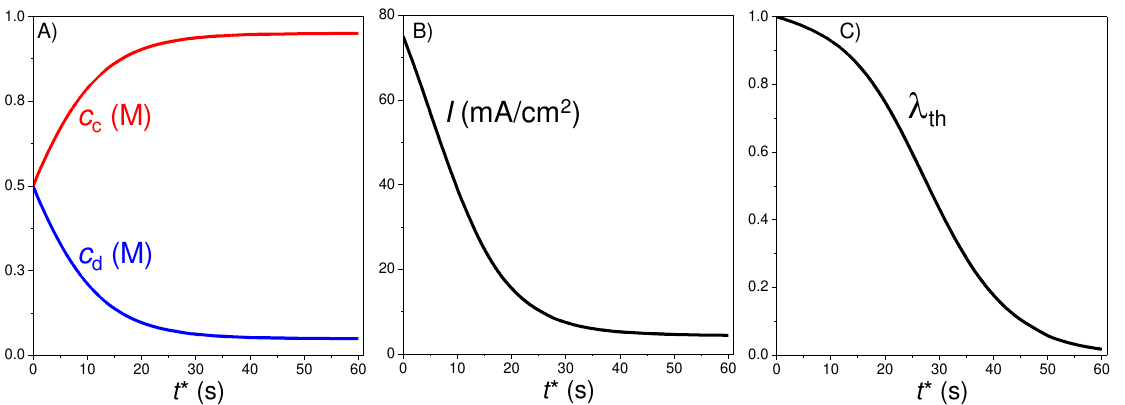}
\vspace{-3mm}
\caption{Calculations of desalination in electrodialysis with co-current flow in diluate and concentrate channels for $V\s{cp}\!=\!180$ mV, $c\s{f}\!=\!0.5$ M, $\text{\WR}\!=\!0.5$, $X\!=\pm\!4.0$ M, $k_\text{m}^*\!=\!1$~$\mu$m/s, $\Phi_i\!=\!1$, and $k\s{ch}\!=\!5$~$\mu$m/s. As function of time-on-stream, \textit{t}\textsuperscript{*}, salt concentrations in the two channels change, first rapidly, then slowly, while current \textit{I} goes down as well as current efficiency, $\lambda\s{th}$. For a high \textit{t}\textsuperscript{*}, there is still a flow of current across the channels and membranes, but the net salt transport is zero, thus $\lambda\s{th}\rightarrow 0$.} 
\label{fig_ED_cocurrent}
\end{figure}

Calculations with this model 
show that while water in the d-channel is desalinated, the Donnan potential strongly increases on the diluate side (by a value which exceeds the decrease on the concentrate side), the current goes down, and current efficiency decreases, finally to reach zero, see Fig.~\ref{fig_ED_cocurrent}. 
In summary, the co-current ED model is a powerful and insightful model to discuss and quantify several well-known observations in ED, such as the decrease of the current density and current efficiency with time, see Fig.~\ref{fig_ED_cocurrent}. 

\begin{oframed}
\label{box_alternative_ion_fluxes_ED}
\underline{Explanation of coion leakage in ED}. Let us describe the functioning of the ED cell pair in an alternative manner. With the current directed to the right, cations will flow rightward, and anions will flow leftward. In the bulk of the channels, for a symmetric system, this current is equally carried by cations, moving right, and anions that move left. Cations and anions can have velocities in this left-right direction that tend to zero (near a blocking membrane), but they do not flow `back' at any point in the module. With AEMs and CEMs alternatingly placed in the stack, and with cations moving right, anions left, then the following will happen. One type of channel has a CEM on its right side. That membrane freely allows the cations to pass to the next channel (to the right). And that same channel has an AEM on its left side, and anions can pass that membrane freely and move left. But now, in these adjacent channels, things are different. The anions want to continue moving left, but now they encounter a CEM which blocks them. And cations that want to move right, encounter an AEM that blocks them. So in this second set of channels, the ions accumulate, and thus the salt concentration increases. These channels will be the concentrate channels. Ions that arrive in these channels, cannot escape. 

What was just described, is the ideal functioning of an ED stack, and it is generally like this early on in a cell pair, where current densities are still high and concentration differences between the two types of channels, $c\s{d}$ and $c\s{c}$, still low.\textsuperscript{1} But further on in the stack, with $c\s{c} \gg c\s{d}$, ions start to leak from the concentrate channel. They leak in the direction that drives them onward, not back to the diluate channel from which they came, but on to the \textit{next} diluate channel. The membrane that blocked them from leaving the concentrate channel in the ideal case (e.g., the AEM for a cation), becomes leaky. Why is that? Chemically the membrane is of course the same as before. But this increased coion leakage from the concentrate channels is because the increasing concentration difference across this membrane leads to more and more diffusion of coions. Indeed, diffusion is the main membrane transport mechanism for co-ions in ED. Thus further on in a cell pair, the driving force for diffusion increases, and thus ions that were kept in the concentrate channel quite well at first, 
now increasingly leak out; so cations that entered a concentrate channel on the left side, now leak out on the right side.

\end{oframed}

\section{Fundamentals of ion and water transport across membranes}
\label{section_fundamentals}

\subsection{Introduction}
In the previous sections we made use of simplified approaches for the transport of solutes and water through RO and ED membranes. In the RO-calculation we neglected charge effects, while in the ED-model we neglected convection of ions. In a generalized approach all these driving forces and frictions are jointly considered, in both types of processes. Also reactions between ions, especially those involving the \ce{H+}-ion in (de-)protonation reactions, are important to include. On both membrane/solution interfaces there is equilibrium (across the interface) both with regard to pressure (mechanical equilibrium), and with regard to chemical potential of solutes (chemical equilibrium). 

For transport across the membrane, be it for membranes as thin as 100 nm as in RO, or membranes 100 to 1000 times thicker for ED, we use continuum theory for flow of water and ions 
applied to transport in the pores in RO and ED membranes. The space charge (SC) model includes profiles in concentration and potential across the cross-section of a pore in the membrane, and also includes the effect of pressure on both membrane interfaces~\cite{Sonin_1976,Peters_2016,Schlogl_1955,Verbrugge_1990,Fair_1971, GuzmanGarcia_1990,Basu_1997,Yang_2004}. But we use a simplification of the SC-model where concentrations are averaged over the cross-section of  pore, which is the Uniform Potential (UP) model. The UP-model is an extension of the Teorell-Meyer-Sievers (TMS) model that describes ion transport across membranes, but TMS theory does not include water flow~\cite{Teorell_1956}. The UP-model can now be extended to incorporate many more effects than we could in the SC model. Besides water-membrane and water-ion friction, we can now also include ion-membrane friction and ion-ion friction, the latter describing how ions that overtake slower ones drag these along, and vice-versa slow ions will retard faster ones. This is the solution-friction (SF) model, and it describes both the flow of water and of solutes~\cite{Tedesco_2017,Oren_2018}. 
In an extended version, also ion volume effects can be included, and this is called the two-fluid model (TFM)~\cite{Kuipers_1992,Biesheuvel_2011,Biesheuvel_Dykstra_2020}. 

In SF theory, the velocity of solutes (such as ions) is determined by a balance between driving forces and frictions. Driving force acting on a solute are mathematically formulated as the negative of a gradient of an ion's chemical potential. Frictions experienced by a solute are with the water flowing through the pores, and with the membrane matrix, which acts as an immobile background structure. For the water a similar force balance is set up but driving forces acting on the water are hydrostatic and osmotic pressures gradients~\cite{Katchalsky_1965,Bacchin_2017}. Thus ions and other solutes are described in a very different manner from how the water that flows past the ions through the pores is described. This is one of the key points of SF theory~\cite{Wang_Sci_Adv_2023,Heiranian_2023,Fan_2024}. 
%
%
This difference in treatment is required because water is a continuum fluid and fills up all available space in the porous medium (such as a membrane), and thus forms continuous liquid pathways. Ions and other solutes are discrete, dispersed entities, and they travel across these pathways of water. These different modes of transport for water and for solutes are essential elements of SF theory.

We will demonstrate the accuracy of SF theory by comparing with results of an experiment where water and solutes flow in opposite directions through an ion-exchange membrane. This osmosis experiment is very hard to describe precisely so it is a very good test of the validity of a transport model. But first we present the fundamental basis of flow of ions (and other solutes) and water according to SF theory. 

\subsection{Flow of solutes through membranes according to solution-friction theory} 
\label{section-RO-ions}

To describe the flow of ions through a porous tortuous medium, first of all we set up an expression for the chemical potential of an ion. In this section the porous medium, or material, is a membrane, so these terms are used interchangeably; we also use the term matrix for the solid phase of the membrane. We can consider multiple contributions to the chemical potential of an ion (unit J/mol) in a porous medium, adding up to~\cite{Biesheuvel_arxiv_2022}
\begin{equation}
\overline{\mu}_i = \overline{\mu}_{\text{ref},i} + RT  \ln \left( c_i / c\s{ref} \right)  + z_i F V  + \overline{\mu}_{\text{aff},i} + \overline{\mu}_{\text{exc},i} + \dots 
\label{eq_fund_part_1}
\end{equation}
where the reference value $\overline{\mu}_{\text{ref},i}$ is of importance when ions take part in chemical reactions, while $RT \, \ln \left( c_i / c\s{ref} \right)$ relates to diffusion (ion entropy), with $c\s{ref}$ a reference concentration, for instance, $c\s{ref}=1$~mM. Furthermore, $z_i F V$ relates to electrostatic effects, $\overline{\mu}_{\text{aff},i}$ to a chemical affinity of the ions with the material, and $\overline{\mu}_{\text{exc},i}$ describes volumetric interactions between ions, and between ions and the porous medium~\cite{Singh_2021}. 
%
%
When only interactions of an ion with the porous medium are of importance (and not between ions), then $\overline{\mu}_{\text{exc},i}$ is a constant factor that can be combined with $\overline{\mu}_{\text{aff},i}$. 

We set up Eq.~\eqref{eq_fund_part_1} for an ion inside and outside the membrane, equate the two expressions, and for a neutral solute then 
arrive for the distribution of a solute across the solution/membrane interface at
\begin{equation}
\frac{c_{\text{m},i }}{ c_{\infty,i}}  = \exp \left( - \tfrac{1}{RT}\, \Delta \left( \overline{\mu}_{\text{aff},i} + \overline{\mu}_{\text{exc},i}\right) \right) = \Phi_i 
\label{eq_fund_part_1a}
\end{equation}
where the partition coefficient, $\Phi_i$, is a constant factor if the chemical affinity and excess terms do not depend (too much) on concentrations in solution and in the membrane, and where $\Delta$ refers to a difference between in the membrane and in solution~\cite{Wang_Biesheuvel_arxiv_2024}. 
The excess contribution is often described as function of an ion size-pore size ratio, $\lambda$, described for a perfectly cylindrical pore and spherical ions by $\Phi_i=\left(1-\lambda_i \right)^2$, but more accurate and realistic expressions are also available~
\cite{Biesheuvel_Dykstra_2020}. For ions we must extend Eq.~\eqref{eq_fund_part_1a} to account for their charge, and we then end up with the extended Boltzmann equation, Eq.~\eqref{eq_Donnan_1}. 


In SF theory we describe the driving forces acting on an ion as minus the gradient of chemical potential. And these driving forces are compensated by all frictional forces, according to a force balance given by
\begin{equation}
\mathcal{F}_{\text{driving},i} + \mathcal{F}_{\text{friction},i} = 0 \, .
\end{equation}
Based on Eq.~\eqref{eq_fund_part_1}, we obtain for the driving force acting on a mole of ions 
\begin{equation}
\mathcal{F}_{\text{driving},i} = - \frac{\partial \overline{\mu}_i}{\partial x} = - RT \left( \frac{1}{c_i} \frac{\partial c_i}{\partial x} + z_i \frac{\partial \phi}{\partial x}  \right)
\end{equation}
where we consider a single coordinate, \textit{x}. An extension to multiple coordinates or dimensions is straightforward. We can leave out here a gradient-term related to $\overline{\mu}_{\text{aff},i}$ because this ion-membrane affinity is invariant across the membrane. And for the same reason we also leave out the excess-term that relates to volume effects (unless membrane compaction is a function of position). 
Thus we only have to consider a force related to diffusion, and to the electric field, i.e., electromigration.

Frictional contributions are because of a friction of the ion with other ions that have a different velocity (otherwise there is no friction), with the water, and with the membrane matrix. 
We neglect ion-ion friction now --otherwise see ref.~\cite{Tedesco_2017}-- so the friction of ions is only with the water that also flows through the porous medium, and with the membrane matrix. Each frictional term is the product of a velocity difference and a friction factor. We then arrive at (p.~128 in ref.~\cite{Katchalsky_1965})
\begin{equation}
\mathcal{F}_{\text{friction},i} = - \sum_j f_{i\text{-}j}^* \left(v_i - v_j\right)=  - f_{i\text{-w}}^*  \left(v_i - J\s{w} \right) - f_{i\text{-m}}^*  \; v_i  
\end{equation}
where water velocity is $J\s{w}$ and we included in the last part that the velocity of the membrane is zero, $v\s{m}\!=\!0$. We present here velocities as superficial velocities, per unit 
total membrane area. A more detailed analysis is the two-fluid model (TFM) which starts at interstitial velocities, and includes pore tortuosity, and also includes how ions take up space that is is not available for water
~\cite{Biesheuvel_Dykstra_2020}.

We can write the ion-water frictional coefficient as the inverse of a diffusion coefficient of the ion in the membrane, $D_{\text{m},i}$, which is the Einstein equation, $f_{i\text{-w}}^* = RT/{D_{\text{m},i}}$~(p.~128 in ref.~\cite{Katchalsky_1965}). This diffusion coefficient is lower than in free solution because 
it includes porosity and tortuosity of the membrane pores. We now obtain an extended Nernst-Planck equation for the molar flux of ions inside 
a membrane (Eq.~(7) in ref.~\cite{Starov_1993}) where convection, diffusion, and electromigration, all contribute~\cite{Teorell_1956,Wang_2021}
\begin{equation}
J_i   = K_{\text{f},i} c_{\text{m},i} J\s{w}  - K_{\text{f},i} D_{\text{m},i} \left( \frac{\partial c_{\text{m},i}}{\partial x} +z_i c_{\text{m},i} \frac{\partial \phi}{\partial x} \right)
\label{eq_fund_NP_ext_twice}
\end{equation}
and where we also incorporated that $J_i = c_{\text{m},i}  v_i$. Concentrations here are those inside the membrane pores. 
The friction factor, or hindrance function, $K_{\text{f},i}$, is given by 
$K_{\text{f},i}=\left(1+ f_{i\text{-m}}^* / f_{i\text{-w}}^* \right)^{-1}$, which has a value between 0 and 1, and describes that ions not only have friction with the 
water that directly envelopes them, but also with the membrane matrix. 
%
%
Eq.~\eqref{eq_fund_NP_ext_twice} is a general expression for ion transport in porous media, extended compared to the standard Nernst-Planck equation with convection and with ion-membrane friction, which leads to the factor $K_{\text{f},i}$. 

\subsection{Flow of water through membranes according to solution-friction theory} 
\label{TFTtheory}

We continue with a description of transport of water through the pores of a porous medium. As discussed before, in SF theory the pores are filled with water completely, and ions have no volume, i.e., in the theory they are point charges. 
ref.~\cite{Biesheuvel_Dykstra_2020}. 
We set up a balance of forces acting on a volume element of water in the membrane pores, with the driving force a gradient in total pressure, i.e., $-\partial P^\text{tot} / \partial x$ is the force acting on a volume element of water. 
The water has a velocity $v\s{w}$ and has friction with the membrane structure, $- f_\text{w-m} RT \left( v\s{w} - v\s{m} \right)$, where $v\s{m}\!=\!0$ because the membrane is immobile~(ref.~\cite{Katchalsky_1965}, p.~126), and water has friction with all solutes, and this friction is proportional to the concentration of solutes, which leads to $- \sum_i^\text{ } f_{i\text{-w}}^* c_{\text{m},i}\left(v\s{w} - v_i\right)$. The summation of all these forces is set to zero, resulting in~\cite{Wang_2021}
\begin{equation}
-\frac{\partial P^\text{tot} }{ \partial x} = f_\text{w-m} RT  v\s{w} + \sum_i f_{i\text{-w}}^* c_{\text{m},i}\left(v\s{w} - v_i \right)  \,.
\label{eq_water_flow_fund_1}
\end{equation}
We can solve Eq.~\eqref{eq_water_flow_fund_1} at each position in the membrane, together with the extended Nernst-Planck equation, Eq.~\eqref{eq_fund_NP_ext_twice}. A significant simplification is arrived at if we assume there is no ion-membrane friction, and thus $K_{\text{f},i}\!=\!1$ in the NP equation, Eq.~\eqref{eq_fund_NP_ext_twice}, for all ions. 
Then inserting the NP equation in Eq.~\eqref{eq_water_flow_fund_1} results in
\begin{equation}
-\frac{1}{RT} \left(\frac{\partial P^\text{h}}{\partial x} - \frac{\partial \Pi} {\partial x} \right) = f_\text{w-m} J\s{w} + \sum_i \left(\frac{\partial c_{\text{m},i}}{\partial x} + z_i c_{\text{m},i} \frac{\partial \phi}{\partial x} \right)  
\label{eq_water_flow_fund_2}
\end{equation}
where we implemented that the total pressure, $P^\text{tot}$, is given by $P^\text{tot}=P^\text{h}-\Pi$, and $D_{\text{m},i}=RT/f_{i\text{-w}}^*$. Because in the ideal case that we discuss here the osmotic pressure is the total ion concentration times \textit{RT}, thus $\Pi =  RT \cdot \sum_i c_{\text{m},i}$, Eq.~\eqref{eq_water_flow_fund_2} simplifies to
\begin{equation}
-\frac{1}{RT}\frac{\partial P^\text{h}}{\partial x} = f_\text{w-m}  J\s{w} - X  \, \frac{\partial \phi}{\partial x}  
\label{eq_water_flow_fund_3}
\end{equation}
and thus we have an exact cancellation of the osmotic pressure acting on the water, and the diffusional forces acting on the ions, which transfer to the water via ion-water friction. We changed notation here for the water velocity, from $v\s{w}$ to $J\s{w}$. But this only occurs when the ions have no friction with other phases other than the water. In the derivation of Eq.~\eqref{eq_water_flow_fund_3} we also included local electroneutrality in the membrane, $\sum_i z_i c_{\text{m},i} + X = 0$. We can rewrite Eq.~\eqref{eq_water_flow_fund_3} to~\cite{Chmiel_2005,Schlogl_1955}
\begin{equation}
J\s{w} = - k_\text{w-m} \left(\frac{1}{RT} \frac{\partial P^\text{h}}{\partial \overline{x}} - X \, \frac{\partial \phi}{\partial \overline{x}}  \right)
\label{eq_water_flow_fund_4}
\end{equation}
and thus water flow is because of a hydrostatic pressure gradient, water-membrane friction, and an electrostatic body force term, which entered this force balance via water-ion interaction. For an uncharged membrane, $X\!=\!0$, and then Eq.~\eqref{eq_water_flow_fund_4} simplifies to the Darcy equation, Eq.~\eqref{eq_RO_waterflow_1a}. Thus water flows across an uncharged membrane not by diffusion but because of an internal pressure gradient~\cite{Pappenheimer_1953,Mauro_1957,Ray_1960,Manning_1968}. 
%
For a constant membrane charge, it can be easily integrated between the two sides of the membrane. Let it be reminded that the last three equations all assumed $K_{\text{f},i}\!=\!1$.

The final element of transport theory through a porous structure filled with a continuum fluid such as water, with dissolved solutes, is mechanical equilibrium at the 
interface between the porous medium and the outside solution. This was already discussed in section~\ref{section_RO}. Here we derive the mechanical equilibrium condition on the basis on Eq.~\eqref{eq_water_flow_fund_2}, because with all gradient terms much larger in a thin interfacial region than the terms proportional to velocities (or in other words, the structure of the interfacial region is not significantly modified by flows of ions and water across it), we obtain the result that the total pressure $P^\text{tot}=P^\text{h} - \Pi $ is invariant across any thin interface~\cite{Sonin_1976,Schlogl_1955,Fair_1971}, and this leads to the earlier conclusion that the change in osmotic pressure $\Pi$ (either up or down) equals the change in hydrostatic pressure $P^\text{h}$ across an interface. 

This analysis now allows us to address the question why water has a tendency to flow to that side of a membrane that has the highest salt concentration in the outside solution. This is because of the two osmotic pressure changes across the two interfaces, and they are different if the outside solutions are different~\cite{Mauro_1957,Mauro_1965}. And both for the case of neutral solutes and an uncharged membrane, and for the case with ions and a charged membrane, this leads to a hydrostatic pressure profile in the membrane such that water flows to the concentrated side~\cite{Ray_1960}. For a neutral membrane the osmotic pressure goes down the most on the high-concentration side, and thus hydrostatic pressure there drops the most as well (from a value outside the membrane). For a highly charged membrane (and when $\Phi_i$ is not too low), osmotic pressure goes up upon entering the membrane. This increase in pressure is the lowest on the high-concentration side. Thus again we develop a pressure profile inside the membrane that pushes the water to the high-salinity side. This flow of water is when the two outside hydrostatic pressures are not too different. Else, by applying a hydrostatic pressure difference across the membrane, we can overcome the osmotic effect, and this is of course what we do in RO and NF. 

\subsection{Counterfluxes of salt and water in osmosis}

\begin{figure}
\centering
\includegraphics[width=0.6\textwidth]{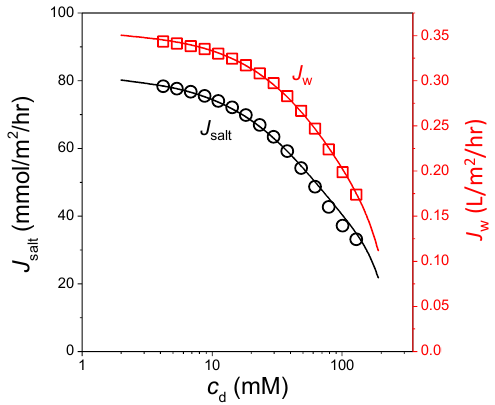}
\vspace{-0mm}
\caption{Experimental and theoretical results of salt flux, $J\s{salt}$, and water velocity, $J\s{w}$, across a cation-exchange membrane between two solutions with different salt concentration (initially $c_\s{d}\!=\!4$~mM and $c_\s{c}\!=\!800$~mM). Water flows to the high-salinity side, and salt in the other direction. There is no external pressure difference and no current. See main text for parameter settings.
}
\label{fig_osmosis}
\end{figure}

To illustrate SF theory, and to show how well it works, we discuss results of an accurate fit of SF theory to data of the flow of water and salt across an ion-exchange membrane. This experiment is called osmosis, and water and salt flow because of a difference in salt concentration between the two outside solutions on either side of the membrane, without pressure differences or current. 
Results of this experiment are presented in Fig.~\ref{fig_osmosis} as function of the concentration on the low-salinity side, $c\s{d}$, which changes in time from 4 mM to 130 mM. During that time the salt concentration on the concentrate side decreases from 800 to 525 mM. Over the 35 hr of the experiment, the volume on the diluate side drops by 40\% because water flows to the concentrated side~\cite{Biesheuvel_Dykstra_2020}. Thus, in this experiment water flows to the high-salinity side, while salt flows in the other direction. Consequently, inside the membrane these two flows are in opposite direction. This makes this an experiment that is not easy to reproduce accurately in a theoretical model, i.e., it is a very good test of a mass transport theory. 

When we apply SF theory, with cation and anion fluxes the same (which is because the electrical current through the membrane is zero), and with equal diffusion coefficients of the ions in the membrane, we find that it works excellently, and results in the very good fit to the data shown in Fig.~\ref{fig_osmosis}. The salt flux, $J\s{salt}$, follows from solving Eq.~\eqref{eq_fund_NP_ext_twice} while implementing that the flux of cations is equal to that of anions, and then we integrate. This leads to Eq.~(5) in ref.~\cite{Biesheuvel_desalination_2023} which we solve assuming $c\s{T,m}$ changes linearly, after which we implement Eq.~\eqref{eq_ED_cTm}. For the partition coefficient, we use $\Phi_i\!=\!0.82$~\cite{Galama_2013}. The water flux is calculated from Eq.~\eqref{eq_water_flow_fund_1} which after integration results in Eq.~(S17) in ref.~\cite{Biesheuvel_desalination_2023}. Membrane charge density is $X\!=\!-5.1$~M. 
To fit the model to the data, we derive  $k\s{m} \! = \! 2.25$~LMH (for a membrane thickness of $L\s{m}=120$~$\mu$m, this corresponds to a diffusion coefficient in the membrane about 20$\times$ less than in free solution), $K_{\text{f},i}\!=\!0.50$, and for the water permeability we use $A \! = \!  9.2$~mLMH/bar. Values for $k\s{m}$ and $A$ are in line with earlier reported values for ED membranes~\cite{Tedesco_2017}. 
Thus, we can conclude that SF theory, which includes chemical and mechanical equilibrium on solution/membrane interfaces, is very well capable of describing fluxes of water and salt in the osmosis experiment. Note again that these fluxes were in opposite directions inside the membrane. This analysis can likely be further improved when we include the activity of ions outside the membrane. Inside the membrane, they are rather constant, thus in effect included in $\Phi_i$~\cite{Wang_Biesheuvel_arxiv_2024}.

It is interesting to estimate the velocities of water and ions in the membrane. We assume a porosity of 34\% and consider velocities straight across the membrane. At the start of the experiment the concentrations on the two sides are 4 mM and 800 mM. The water velocity is $\sim \! 300$ nm/s. 
The counterion velocity is quite constant across the membrane at $\sim 15$~
nm/s. This is a velocity in the direction opposite to the water flow, about 5\% of the water velocity. Thus the counterions have a velocity relative to the water to not be dragged along, and then a 5\% extra velocity to make them move in the reference frame of the membrane. The coions have a concentration that on the high-salinity side is about a factor of 100 less than the counterions, and many more orders smaller on the low-salinity side. If velocity is calculated as a flux divided by a concentration, one finds that the velocity of the coions is much higher than of counterions, and rapidly increases towards the low-salinity side. But if coions are predominantly transported by diffusion, this interpretation of velocity must not be taken too literally -- it is not the case that an individual coion accelerates while moving across the membrane~\cite{Brogioli_2000}. 

\section{Conclusions and Outlook}
\label{section_conclusions}

In this tutorial review, we presented theory for reverse osmosis (RO) and electrodialysis (ED), explaining how both technologies are based on the same fundamental transport theory. 
This is the solution-friction (SF) theory, and for ED we solved it in the absence of convection, thus we did not discuss pressures. We used SF theory for RO but then only described neutral solutes. Finally we solved SF theory for the osmosis experiment based on ions and a charged membrane, and we compared with experimental data. For ED we also developed new equations for Donnan equilibrium that extend the standard ideal expression. We present analytical equations for current efficiency, showing that this is a process parameter, not a membrane material property. For RO we summarized literature for SF theory for neutral solutes including also the effect of concentration polarization. The general derivation we provided of SF theory also results in the twice-extended Nernst-Planck equation which is generally applicable in describing ion flow in reverse osmosis and nanofiltration of salt solutions.

Topics that we did not address in this tutorial review are first of all that both for RO and NF we must implement the Nernst-Planck equation for ions and a charged membrane in a full module calculation, and beyond that extend the theory from simple 1:1 salt solutions to multi-ionic solutions, as we also should for electrodialysis. Even the addition of one extra type of anion or cation can significantly change the entire modeling framework. In addition, in real water sources also the protonation degree of ions must be considered, which depends on local pH. At high concentrations, ions also associate in ion pairs. These effects are relevant to study because for instance an ammonium ion is acted on by the electrical field, but the neutral ammonia species is not. Thus rejection of these ions is strongly pH-dependent. For very tiny pores, a related topic is the effective size of ions, that has an impact on their partitioning and their mobility within the membranes. Ions with a higher charge will be hydrated better, and are expected to be slower. State-of-the-art theory for simultaneous transport and reaction of ions (such as acid-base reactions between ions) assumes that these reactions are very fast, but it is interesting to investigate whether that is a correct assumption. Another important assumption is local electroneutrality in channels and in membranes. Especially in reverse osmosis with membranes as thin as 100 nm, it is important to know if possibly Poisson's equation must be used to replace the assumption of local electroneutrality in the membrane. 

For electrodialysis we also have to include the effect of changes of concentration across the channel width (from membrane to membrane) and for RO and salt solutions, we must do this as well~\cite{Biesheuvel_arxiv_2024_CP}. Other topics of relevance are the use of NF and RO to remove organic micropollutants (OMPs) and other charged molecules. 
A very different topic is the theoretical study of electro-deionization (EDI), which is an ED system with channels filled with (mixed bed) resin particles to reduce energy costs and produce ultra-pure water. In the theory for EDI it may be necessary to include \ce{H+} and \ce{OH-} formation at points where acidic and basic resin particles touch. 
%
%
%
%
%
%
%
%
%
%
%

\section*{Nomenclature}
\begin{small}	
	\begin{longtable}{|l|l|l|}
		\hline 
		\multicolumn{3}{|c|}{Symbols} \\
		\hline 
		$a$ & specific membrane surface area, $a=1/L_\mathrm{ch}$ & m\textsuperscript{-1} \\
		$c_{i,j}$ & concentration of solute or ion $i$ in stream $j$ (*) & mol/m\textsuperscript{3} \\
		$c_{\mathrm{m},i}$ & concentration of ion $i$ in membrane & mol/m\textsuperscript{3} \\
		$c_\mathrm{T,m}$ & total concentration of all ions together 
		in the membrane & mol/m\textsuperscript{3} \\
		$D_{\mathrm{m},i}$ & membrane-based diffusion coefficient, $D_{\text{m},i}=\varepsilon \, D_{\infty,i}$ & m\textsuperscript{2}/s \\
		$e_\text{min}$ & theoretical minimum energy of a desalination process & W \\
		$E_\text{min}$ & theoretical minimum energy of desalination per m\textsuperscript{3} of freshwater produced & J/m\textsuperscript{3} \\
		$F$ & Faraday's constant (96485 C/mol) & C/mol  \\
		$\mathcal{F}_{\text{driving},i}$ & driving force on ion $i$ & J/mol/m\\
		$\mathcal{F}_{\text{friction},i}$ & friction force on ion $i$ & J/mol/m \\
		$f_{\mathrm{cou},j}$ & free energy density of Coulombic interactions between ions 
		& J/m\textsuperscript{3} \\
		$f_{\text{w-m}}$& friction factor of water with the membrane matrix & mol.s/m\textsuperscript{5} \\ 
		$f_{i\text{-}j}^*$ & friction factor of ion $i$ with another ion $j$ & J.s/mol/m\textsuperscript{2} \\
		$f_{i\text{-m}}^*$& friction factor of an ion with the membrane matrix & J.s/mol/m\textsuperscript{2} \\ 
		$f_{i\text{-w}}^*$ & friction factor of an ion with water & J.s/mol/m\textsuperscript{2} \\
		$I$ & current, or current density & A, or A/m\textsuperscript{2} \\
		$J_i$ & transmembrane molar flow rate of solutes (**) & mol/m\textsuperscript{2}/s \\
		$k_\mathrm{ch}$ & channel mass transfer coefficient, $k_\mathrm{ch}=\varepsilon D_\infty/L_\mathrm{ch}$ & m/s \\
		$k_\text{w-m}$ & water-membrane permeability, $= 1 / \left(f_\text{w-m} L\s{m} \right)$ & m\textsuperscript{4}/mol.s \\
		$k_{\mathrm{m},i}$ & membrane mass transfer coefficient, $k_{\mathrm{m},i}=D_{\mathrm{m},i}/L_\mathrm{m}$ & m/s \\
  		$k^*_{\mathrm{m},i}$ & modified membrane mass transfer coefficient, $k^*_{\mathrm{m},i}=K_{\text{f},i} k_{\mathrm{m},i}$ & m/s \\
		$K_{\mathrm{f},i}$ & function describing solute-membrane friction & \\
		$L_\mathrm{ch}$ & flow channel thickness & m \\
		$L_\mathrm{m}$ & membrane thickness & m \\
		
		$p$ & membrane porosity &  \\		
		$P^\mathrm{h}$ & hydrostatic pressure & Pa (bar) \\
		$P_i$ & passage of ion or solute $i$, $P_i = 1-R_i$ &  \\ 
		$\mathrm{Pe}_i$ & membrane P\'{e}clet-number, $\mathrm{Pe}_i=v_{\mathrm{w}}/k_{\mathrm{m},i}$ &  \\
		$R$ & Gas constant (8.3144 J/mol/K) & J/mol/K  \\
		$R_i$ & rejection (or retention) of ion or solute $i$ &  \\
		$S_j$ & entropy associated with a certain water flow & J/s/K \\
		$T$ & temperature & K \\
		$t^{*}$ & time-on-stream, $t^{*}=z/v_z$ & s \\
		$V_\mathrm{cp}$ & cell pair voltage & V \\
		$v_{\mathrm{w}}$ & transmembrane water velocity, permeate water flux (**) & m/s \\
		$V_\mathrm{T}$ & thermal voltage, $V_\mathrm{T}=RT/F$ & V \\
		$v_z$ & crossflow velocity of water through a channel, i.e., along membrane & m/s \\
		\WR& water recovery &  \\
		$x$ & coordinate towards and across membrane &  \\
		$\left| X \right|$ & magnitude of the membrane charge density & mol/m\textsuperscript{3} \\
		$z$ & direction along membrane &  \\
        $z_i$ & ion valency & \\
		$\Delta P^{\mathrm{h},\infty}$ & hydrostatic pressure difference across the membrane & Pa (bar) \\
		$\Delta \phi_{\mathrm{ch},j}$ & potential drop across channel $j$ (dimensionless) &  \\
		$\Delta \phi_{\mathrm{D},j}$ & Donnan potential at solution/membrane interface &  \\
		$\Delta \phi_{\mathrm{D,tot}}$ & total Donnan potential across a membrane &  \\
		$\Delta \phi_\mathrm{m}$ & potential drop across the inner coordinates of the membrane & \\
		$\varepsilon$ & $\varepsilon = p/\bm{\tau} $  &  \\
		$\lambda$ & current efficiency (index `cp' for cell pair, and index `th' for theoretical) & \\
		$\mu_i$ & chemical potential of an ion & J/mol \\
		$\mu_{\mathrm{aff},i}$ & ion chemical potential due to affinity  & J/mol \\
		$\mu_{\mathrm{exc},i}$ & ion chemical potential due to volumetric interactions & J/mol \\
		$\nu_\mathrm{i}$ & molar volume of an ion  & m\textsuperscript{3}/mol\\
		$\Pi$ & osmotic pressure of a solution & Pa (bar) \\
		$\Pi_\mathrm{int}$ & osmotic pressure at the interface between CP layer and membrane & Pa (bar) \\
		$\sigma_i$ & reflection coefficient, $\sigma_i=1-K_{\text{f},i} \Phi_i$ & \\
		$\tau$ & tortuosity & \\
		$\bm{\tau}$ & tortuosity factor of a porous structure $\bm{\tau}= \tau^2$ & \\
        $\phi$ & dimensionless electrical potential & \\
		$\phi$ & volume fraction of all ions together  & \\
		$\phi_{\text{v}}$ & volumetric flowrate & m\textsuperscript{3}/s \\
		$\Phi_i$ & partition coefficient \\ 
        $\omega$ & transport factor in RO of neutral solutes & \\
		
		\hline 
		\multicolumn{3}{|c|}{Subscripts} \\
		\hline 
		$\text{c}$ & concentrate (in RO: retentate) &  \\
		$\text{f}$ & feedwater  & \\
		$\text{p}$ & product water, or freshwater (in RO: permeate) &   \\
		$\infty$ & conditions outside membrane; salt concentration in solution for $z\!:\!z$ salt &  \\
		\hline
	\end{longtable}

\noindent
(*) The words ion and solute are interchangeably used in this table. \\
(**) These fluxes across the membrane are defined per unit geometrical, i.e., outer, area of a membrane, i.e., they are defined as superficial velocities, not interstitial. This means they are not defined for instance per cross-sectional area of the pores only.

\end{small}

\end{document}